\documentclass[11pt]{article}

\usepackage[margin=2.5cm]{geometry}
\usepackage{amsmath,amssymb}
\usepackage{microtype}
\usepackage{revtexauthors}
\usepackage{float}
\usepackage{graphicx}
\usepackage{hyperref}
\usepackage{dsfont}
\usepackage{centernot}
\usepackage[normalem]{ulem}
\usepackage{xcolor}

\graphicspath{{img/}}

\newcommand{\ket}[1]{|#1\rangle}
\newcommand{\bra}[1]{\langle #1|}
\newcommand{\sket}[1]{|#1]}
\newcommand{\sbra}[1]{[ #1|}

\newcommand{\braket}[2]{\langle #1 \vert #2 \rangle}
\newcommand{\sbraket}[2]{[ #1 \vert #2 \rangle}
\newcommand{\brasket}[2]{\langle #1 \vert #2 ]}
\newcommand{\sbrasket}[2]{[ #1 \vert #2 ]}

\newcommand{\Tr}{\mathrm{Tr}}

\newcommand{\SU}{\mathrm{SU}}
\newcommand{\U}{\mathrm{U}}
\newcommand{\SL}{\mathrm{SL}}

\newcommand{\R}{\mathbb{R}}
\newcommand{\C}{\mathbb{C}}
\newcommand{\Id}{\mathds{1}}

\renewcommand{\Re}{\mathrm{Re}}
\renewcommand{\Im}{\mathrm{Im}}

\renewcommand{\d}{\mathrm{d}}

\newcommand{\EPRLclosed}{\text{\tiny EPRL-FK-closed}}
\newcommand{\EPRLwedge}{\text{\tiny EPRL-FK-wedge}}

\title{Closing the Loop: from EPRL-FK spinfoams to Regge dynamics}

\author{Matteo Bruno}
\email{matteo.bruno@cpt.univ-mrs.fr}
\affiliation{Aix-Marseille Univ, Universit\'e de Toulon, CNRS, CPT, Marseille, France}

\author{Pietro Don\`a}
\email{pietro.dona@cpt.univ-mrs.fr}
\affiliation{Aix-Marseille Univ, Universit\'e de Toulon, CNRS, CPT, Marseille, France}

\author{Gowrisankar Sreeram}
\email{gowrisankar.sreeram@cpt.univ-mrs.fr}
\affiliation{Aix-Marseille Univ, Universit\'e de Toulon, CNRS, CPT, Marseille, France}

\date{\today}

\begin{document}

\maketitle

\begin{abstract}
We study the semiclassical limit of the EPRL-FK model on a fixed simplicial complex. We ask what additional condition is needed for its variational principle to reproduce the equations of motion of Regge calculus, the standard discretization of General Relativity. This limit differs and complements the definition of the semiclassical regime in which, for suitable boundary states and a sufficiently refined simplicial complex, spinfoam amplitudes approximate gravitational transition amplitudes to a prescribed accuracy. We recast the EPRL-FK model in wedge-holonomy variables and identify geometric closure as the missing condition. We prove that imposing geometric closure as an additional constraint alongside local flatness allows us to reconstruct curved geometries. On a regular, nondegenerate Lorentzian branch with spacelike tetrahedra, these constraints define a surface equivalent to the space of length-Regge geometries, modulo gauge. The wedge action restricted to this surface becomes the Regge action, and variations tangent to this surface yield the length-Regge equations. We propose a way to implement geometric closure directly in the spinfoam amplitude. With this additional constraint, the EPRL-FK model is equivalent to Regge calculus in its semiclassical regime on a fixed simplicial complex.
\end{abstract}

\section{Introduction}
\label{sec:intro}

Spinfoams provide a covariant description of the dynamics of Loop Quantum Gravity \cite{Rovelli:2014ssa}. The kinematical states of the theory are spin-network states, which encode three-dimensional quantum geometries on a boundary graph. A spinfoam assigns a transition amplitude between such boundary states by summing over histories interpolating between them. These histories are two-complexes, labelled by spins, intertwiners, and group variables. In this sense, a spinfoam model plays the role of a path integral for Loop Quantum Gravity. Its classical limit is expected to reproduce General Relativity, or an appropriate discrete approximation to it.

Simply increasing the scale of the boundary geometry is not enough to achieve a prescribed semiclassical accuracy: the discretization must also be sufficiently refined \cite{Engle:2021xfs}. Taking the asymptotic limit on a fixed simplicial complex is genuinely different from making a semiclassical approximation at large but finite spins, with a discretization that is adapted to the geometry. This distinction belongs to a broader discussion of how Regge dynamics emerges from spinfoams, including earlier approaches combining large areas with a small Immirzi parameter \cite{Magliaro:2011dz,Han:2013ina}.

Quantitative support for such a semiclassical regime comes from analyses combining large spins and small curvature, both in standard and effective spinfoam models \cite{Han:2013tap,Asante:2020qpa,Asante:2020iwm}. In this regime, the simplices must be simultaneously large compared to the Planck scale and small compared to the curvature scale. These considerations suggest selecting a finite semiclassical accuracy first and then choosing a discretization fine enough to resolve the curvature while keeping each simplex within the semiclassical regime. Related work has investigated the recovery of Regge equations and a small-curvature window from a regulated spin sum, together with a coupled semiclassical and refinement limit \cite{Han:2017xwo}.

Our work is complementary to these approaches: we keep the simplicial complex fixed and ask how Regge calculus, the standard discretization of General Relativity, could be encoded in the semiclassical variables of a spinfoam model. More specifically, we seek the constraints on spins and holonomies that make the real constrained semiclassical variational principle reproduce the length-Regge equations of motion.

Physical spinfoam models are commonly constructed from topological BF theory, which carries no local degrees of freedom, by imposing simplicity constraints weakly at the quantum level. The current best-studied realization of this construction is the EPRL-FK model (Engle-Pereira-Rovelli-Livine\textendash Freidel-Krasnov) \cite{Engle:2007wy, Freidel:2007py, Kaminski:2009fm}. This implementation selects gravitational data, but does not by itself establish that the resulting amplitude reproduces the dynamics of General Relativity or its Regge discretization.

The large-spin asymptotics of the Lorentzian EPRL-FK vertex provided an important first step: for suitable nondegenerate geometric boundary data, its critical configurations reconstruct a Regge simplex, and the action evaluated on them gives the corresponding Regge phase \cite{Conrady:2008mk,Barrett:2009mw}. This establishes the local emergence of Regge geometry.

At fixed internal spins, the large-spin analysis on simplicial complexes with many vertices likewise reconstructs Regge geometries on the nondegenerate geometric branch, with a phase given by the Regge action of the complex \cite{Han:2011re, Han:2011rf}. Recovering this phase, however, leaves open the question of which equations follow when the internal spins are also varied. Requiring stationarity with respect to the internal spins leads to the so-called flatness problem: when the spins are treated as continuous, the real stationary equations on the nondegenerate, consistently oriented geometric branch require vanishing deficit angles. This was first noted \cite{Conrady:2008mk}, sharpened \cite{Bonzom:2009hw}, and later given a more precise formulation accounting for the discrete spin sum \cite{Hellmann:2013gva}. Its implications for the semiclassical limit remain debated \cite{Oliveira:2017osu, Engle:2021xfs}. In particular, a restriction on the real stationary configurations at fixed complex need not exclude curved contributions at large but finite spins in the regimes discussed above.

In the $\SU(2)$ BF setting, vertex boundary data were related to polytope geometries and the corresponding actions to $\SU(2)$ graph invariants \cite{Dona:2017dvf}. Numerical tools for EPRL-FK amplitudes \cite{Dona:2018nev} enabled tests of Regge asymptotics at moderate spins \cite{Dona:2019dkf} and of the suppression of curved configurations in the examples studied \cite{Dona:2020tvv}. The local reconstruction of simplex geometry was also traced to local flatness of the wedge holonomies, which requires parallel transport to be trivial around cycles within each vertex \cite{Dona:2022hgr}. This condition permits nonzero curvature around bulk triangles: curvature is encoded by the holonomy around a face traversing several vertices.

Analyses based on an analytic continuation of the integration variables located complex critical points associated with curved Regge geometries with small but nonzero deficit angles \cite{Han:2021kll}. Subsequent works developed an effective action for Regge geometries on complexes with several vertices and found that its dynamics approximates the Regge equations in the small Barbero-Immirzi parameter regime \cite{Han:2023cen}. Numerical tools were also developed to locate complex critical points more broadly \cite{Han:2024lti}. In the small-curvature regime studied at finite Immirzi parameter, unsuppressed curved contributions lie within a window that narrows as the spin scale increases and widens as the Immirzi parameter decreases \cite{Han:2023cen}. The complexified equations can also admit configurations without a Regge interpretation, so this approach does not by itself establish equivalence with length-Regge calculus.

On a family of regular lattices, studies of the continuum dynamics of spinfoam area and angle variables recovered the graviton dynamics of General Relativity at leading order, together with corrections associated with nonmetric degrees of freedom \cite{Dittrich:2022yoo}. A subsequent program constructed candidate continuum actions for these additional degrees of freedom and classified which ones remain free of unphysical ghost modes \cite{Borissova:2022clg, Borissova:2023yxs}. These studies investigate how the larger geometric configuration space affects the continuum dynamics.

Recovering Regge dynamics through an additional geometric constraint on a fixed complex requires specifying both the independent variables and their allowed variations, since identifying a Regge-like phase alone does not determine the dynamics. Earlier formulations explored independent dihedral angles and constrained triangle areas \cite{Barrett:1994nn,Makela:1998wi,Makela:2000ej}. Area-angle Regge calculus subsequently provided a local formulation in which closure and shape-consistency conditions restrict areas and three-dimensional angles to configurations admitting consistent edge lengths \cite{Dittrich:2008va}. Related phase-space analyses clarified how constraints on bivectors and holonomies reduce the larger space of discrete connection data to geometric configurations, distinguishing twisted geometries from the more restricted Regge sector \cite{Dittrich:2008ar,Dittrich:2010ey,Dittrich:2012rj}.

In a Riemannian formulation of spinfoam dynamics, simplicity and parallel-transport relations were shown to encode the consistency conditions of area-angle Regge calculus, with closure implemented as an explicit restriction coupling areas and normals \cite{Bonzom:2009wm}. A later study related the EPRL-FK constructions to path integrals and identified the flatness equation arising from independent area variations \cite{Bonzom:2009hw}. It also proposed imposing closure independently to constrain these variations and recover the expected discrete gravitational dynamics.

The implementation of closure through quantum intertwiners must be distinguished from imposing it as an independent geometric constraint \cite{Alexandrov:2008da}. Path-integral constructions show that the choice of variables and the way constraints are imposed can change the resulting quantum model \cite{Han:2010rb,Baratin:2011hp}. Related work used closure conditions on tetrahedral four-normals to complete the geometric constraints \cite{Gielen:2010cu}. A subsequent Hamiltonian formulation related to EPRL-FK identified a potentially missing torsional condition and proposed conservation of volume-weighted tetrahedral normals at each vertex \cite{Wieland:2013cr}. A later construction promoted these normals to dynamical variables, introduced a volume-dependent mass-shell condition, and obtained configurations with nonzero face curvature \cite{Wieland:2014nka}. Its configuration space still includes twisted geometries, and the construction does not establish equivalence with Regge calculus. The four-dimensional closure of tetrahedral normals around a four-simplex is distinct from the area-weighted closure of triangle normals within a tetrahedron that we study here.

Analyses of area Regge calculus explore nonmetric degrees of freedom by characterizing the configurations permitted when neighboring simplices agree on shared areas but need not agree on their shapes \cite{Asante:2018wqy}. At the amplitude level, formulations with explicit gluing factors investigate how closure and shape-matching violations are suppressed, and how these factors connect the quantum amplitude to descriptions in terms of semiclassical vertices \cite{Asante:2022lnp}. Together, these approaches motivate examining precisely which constraints tie the area and shape data of a spinfoam to a consistent length geometry.

In the EPRL-FK model, wedge holonomies and spins make the relation between spinfoam variables and discrete geometry explicit \cite{Dona:2022hgr}. The holonomies encode the dihedral and shape data, while the spins supply the independent area variables. Local flatness fixes shape matching within each four-simplex and reconstructs its tetrahedral shapes, but does not by itself tie the spins to those shapes. This requires a compatibility condition between the spins and the normals encoded by the wedge holonomies.

This compatibility condition, which we call \emph{geometric closure} or holonomy-flux compatibility, ties the spin areas to the tetrahedral geometry reconstructed from the holonomies. It already holds for geometric critical configurations of the model, but is not imposed independently off shell and therefore does not restrict the allowed spin variations. It is distinct from the quantum Gauss constraint on EPRL-FK intertwiners and from closure of coherent-intertwiner labels. Our contribution formulates this compatibility condition in Lorentzian wedge-holonomy variables and proves its role in recovering Regge dynamics.

Sections~\ref{sec:setup}--\ref{sec:symmetries} formulate the EPRL-FK amplitude in wedge-holonomy variables, recall the role of local flatness, and identify the relevant gauge symmetries. Section~\ref{sec:eom} derives the flatness equation from independent spin variations. Section~\ref{sec:dynamical-vs-strong} explains the difference between closure as an equation satisfied at critical points and closure imposed as an independent constraint. Section~\ref{sec:closure-spinors} proves the equivalence of the resulting constrained variational principle with length-Regge calculus on the regular geometric branch specified above. Finally, Section~\ref{sec:changing-eprl-fk-edge} discusses the limitations of closed coherent-intertwiner resolutions and proposes a candidate implementation of geometric closure in the amplitude.

\section{The EPRL-FK spinfoam model and local flatness}
When formulated in terms of wedge holonomies and supplemented with local-flatness constraints, the EPRL-FK model admits a direct interpretation in terms of simplicial geometries and makes the semiclassical geometry transparent. Local flatness constrains the holonomy data and imposes shape matching among the tetrahedra, thereby reconstructing a Lorentzian $4$-simplex at each vertex. However, it does not require the spins $j_f$, interpreted as triangle areas, to be compatible with the reconstructed shapes. At the level of the action, the area and shape degrees of freedom therefore remain independent. As we explain in the final subsection, geometric closure relates them on shell but is not imposed as an independent off-shell constraint. Consequently, the spins can be varied freely, leading to vanishing deficit angles rather than to the Regge equations. We will provide the details of this formulation in this section.
\subsection{The EPRL-FK spinfoam model in terms of wedge holonomies}
\label{sec:setup}
\begin{figure}[H]
    \centering
    \includegraphics[scale=1]{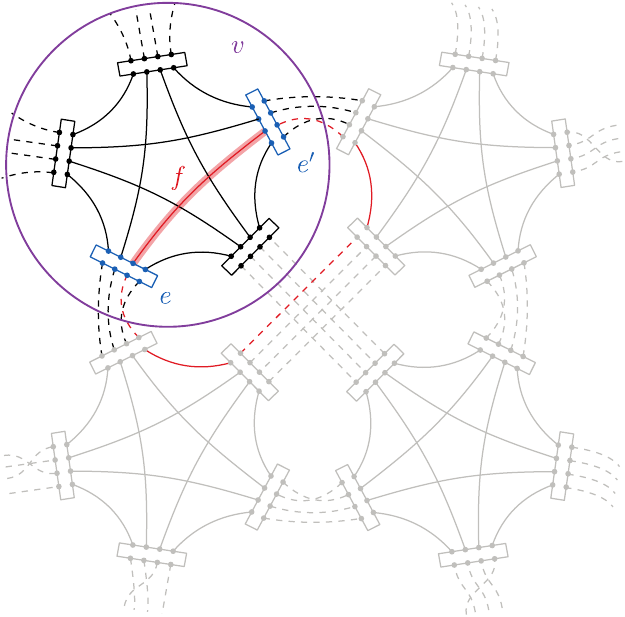}
    \caption{\label{fig:2complex} Zoom on the wiring diagram of a two-complex $\Delta$: four vertices are shown, with one vertex, $v$, circled in purple. The face $f$, in red, belongs to $v$ but is also shared with two further vertices. At $v$, the face $f$ identifies the wedge $vf$, highlighted in red. The source and target of the wedge $vf$ are the edges $e$ and $e'$, respectively, shown in blue. The orientation of the face is not depicted.}
\end{figure}

First, we revisit known material from \cite{Dona:2022hgr} in a modern language, fix the notation used in the rest of the paper, and lay the foundations for what follows. The EPRL-FK spinfoam amplitude \cite{Engle:2007wy, Freidel:2007py} on a two-complex $\Delta$, with faces colored by $\SU(2)$ spins $j_f$ and edges colored by $\SU(2)$ intertwiners $i_e$, is
\begin{equation}
    \label{eq:transition-amplitude}
    A_\Delta = \sum_{j_f, i_e} \prod_f A_f(j_f) \prod_e A_e(i_e) \prod_v A_v(j_f,i_e) \, .
\end{equation}
The standard face and edge amplitudes, $A_f = 2j_f+1$ and $A_e = 2i_e+1$, are fixed by the gluing properties of the spinfoam path integral \cite{Bianchi:2010fj}. The model-dependent structure relevant for our analysis is contained in the vertex amplitude $A_v$, which we decompose into contributions from \emph{wedges}: the portions of a face $f$ lying within a vertex $v$, bounded by the two edges $e,e'$ of $\Delta$ shared by $f$ and $v$. We denote a wedge by the pair $vf$, adding the edges as subscripts, $vf_{ee'}$, when needed. See Figure~\ref{fig:2complex} for a pictorial representation of the elements of a simplicial complex.
Each wedge contributes to $A_v$ with the matrix element of a $\gamma$-simple $\SL(2,\C)$ principal-series representation in the lowest-weight $\SU(2)$ sector of spin $j_f$, with $\gamma$ the Barbero-Immirzi parameter:
\begin{equation}
   vf \quad \longrightarrow  \quad D^{(\gamma j_f, j_f)}(g_{vf}) \, .
\end{equation}
This representation is selected by the EPRL-FK $Y^\gamma$ map, which is the foundation of the model. The map implements the quantum linear simplicity constraint and selects, among all representations, those relevant for reducing a topological theory to gravity.

The group element $g_{vf}$ is the wedge holonomy: we take these, rather than the usual half-edge holonomies, as the fundamental variables of the theory \cite{Dona:2022hgr}. Since each wedge is bounded by two edges, this gives many more holonomies than the standard formulation (10 per vertex vs. 5), and we will need to impose constraints to recover the usual theory. We return to this in Section~\ref{sec:local-flatness}. For now, we focus on how geometry is encoded in the wedge holonomies $g_{vf}$.

\begin{figure}[H]
\includegraphics[scale=0.8]{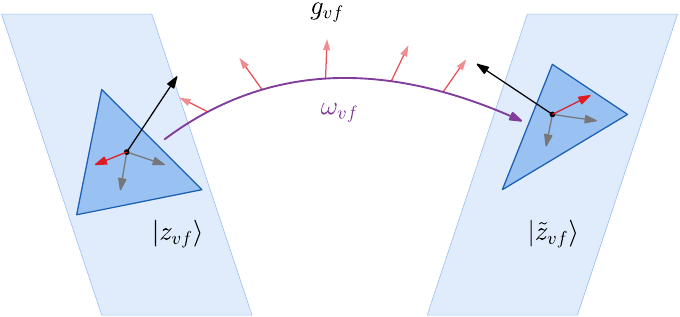}
\centering
\caption{\label{fig:wedge-holonomy} The wedge holonomy $g_{vf}$ as a map between triangles on framed planes. The red vector is shown schematically as it is parallel transported from one framed triangle to the other. The twist angle measures its rotation with respect to the two reference frames, shown as gray arrows.}
\end{figure}

The main advantage of wedge holonomies is their explicit geometric interpretation: we parametrize $g_{vf}$ by two unit-norm spinors and a complex angle (our spinor conventions and the identities used throughout are collected in Appendix~\ref{app:spinors})
\begin{equation}
    \label{eq:wedge-param}
    g_{vf} = e^{\omega_{vf}/2} \sket{\tilde z_{vf}}\bra{z_{vf}} - e^{-\omega_{vf}/2} \ket{\tilde z_{vf}}\sbra{z_{vf}} \, ,
\end{equation}
where $\ket{z_{vf}}$ and $\ket{\tilde z_{vf}}$ are the source and target spinors of the wedge. Square brackets denote the dual spinor, in the notation of twisted geometries \cite{Freidel:2010aq, Freidel:2010tt, Borja:2010rc, Livine:2011gp}. Each spinor determines a three-dimensional reference frame associated with a framed plane: its direction, $\vec n_{vf} = -\bra{z_{vf}}\vec\sigma\ket{z_{vf}}$, is interpreted as the plane's normal\footnote{The modulus of the scalar product between any two unit spinors fixes the angle between their normals, \eqref{eq:app-sp-angle} in Appendix~\ref{app:spinors}, with $\theta$ the dihedral angle between the corresponding faces.}, while its phase singles out a reference direction within the plane. This phase is the extra \emph{framing} needed to compare vectors between the two planes. The holonomy then maps the \emph{framed plane} of the source triangle to that of the target, while $\omega_{vf}$ encodes the boost (real part) and twist (imaginary part) relating the two, as depicted in Figure~\ref{fig:wedge-holonomy}.
Writing the vertex amplitude itself requires introducing some auxiliary spinors. Since they are not part of the model's fundamental data and are eventually integrated out, we defer its explicit form to Appendix~\ref{app:vertex-amplitude-details} where we also include further details of the wedge parametrization of the holonomy. In the large-spin limit, the amplitude is controlled by the stationary points of the exponent appearing in the path integral. We call this exponent the semiclassical action. With this convention the amplitude is written as $\exp S$, and the factor of $i$ multiplying the real phase is included in $S$ itself.
\begin{equation}
    \label{eq:amplitude-exp}
    A_\Delta = \sum_{j_f} \int \prod_{vf} \d g_{vf} \, e^{S^{\text{EPRL-FK}}(j_f, g_{vf})} \, .
\end{equation}
After imposing the alignment and the dominance conditions on the auxiliary variables, the contribution to the action from the wedges is
\begin{equation}
    \label{eq:EPRL-wedge}
    S^{\EPRLwedge} = i\sum_{f} j_{f} \sum_{v\subset f}\left( \gamma \, \Re\,\omega_{vf} + \Im\,\omega_{vf} \right) \, .
\end{equation}
Integrating out the auxiliary spinors imposes an extra condition on the wedge holonomies: the spinors $\ket{z_{vf}}$ and $\sket{\tilde z_{v'f}}$ of two consecutive wedges, a priori independent, both represent the same shared triangle and must therefore coincide, $\ket{z_{vf}} = \sket{\tilde z_{v'f}}$\footnote{Formally they have to coincide only up to a phase. This phase is arbitrary and, as we will see, is part of the redundancy of the parametrization we chose. We fix it to be trivial for simplicity.}. This identification, the \emph{framing} of the triangle, is what makes the wedge holonomy compatible across the edge. Once identified, this common spinor no longer depends on which of the two wedges is used to define it, only on the edge $e$ shared by $v$ and $v'$ and on the face $f$\footnote{With this identification, the composition of two consecutive wedge holonomies, $g_{v'f}g_{vf}$, again has the form \eqref{eq:wedge-param}: its source spinor is that of $g_{vf}$, its target spinor is that of $g_{v'f}$, and its complex angle is the sum of the two wedge angles.}. From now on we drop the vertex label and write it as $\ket{z_{ef}} \equiv \ket{z_{vf}} = \sket{\tilde z_{v'f}}$.

\begin{figure}[H]
\centering
\includegraphics[scale=1]{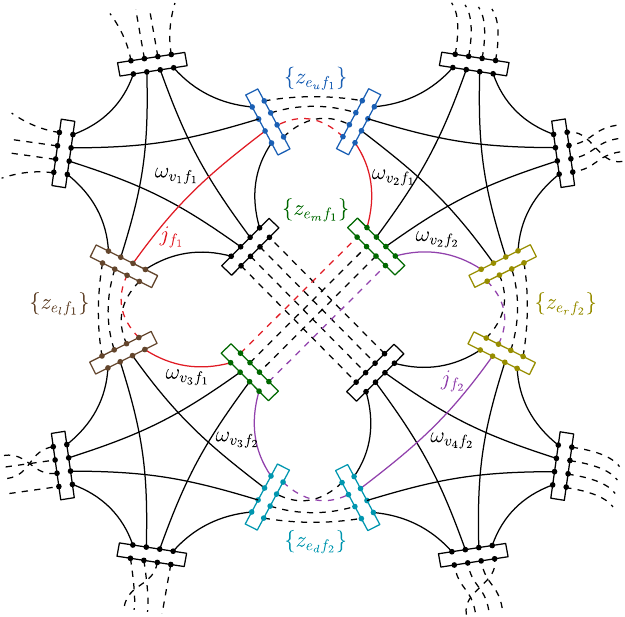}
\caption{\label{fig:notation} Wiring diagram of a two-complex, illustrating the spin and spinor notation. 
Each face carries a spin. We highlight two, of spin $j_{f_1}$ (red) and $j_{f_2}$ (purple). Each edge, dual to a tetrahedron, carries four spinors, one for each of its triangles. We single out five edges, each drawn in its own color: up ($e_u$), right ($e_r$), down ($e_d$), left ($e_l$), and middle ($e_m$). We denote by $\{z_{ef}\}$ the spinor associated to edge $e$ and face $f$.
}
\end{figure}

The holonomy around a face, $g_f = \prod_{v\subset f} g_{vf}$, is a 4-screw (a boost and rotation in the same direction) about the framed plane of the source triangle. The total boost angle is $\sum_{v\subset f}\Re\,\omega_{vf}$ and the total twist angle is $\sum_{v\subset f}\Im\,\omega_{vf}$. At this point the action already resembles $i$ times the Regge action, but with a major difference: it does not yet encode any four-dimensional geometry.

The fundamental variables of the model are the spins $j_f$ and the holonomies $g_{vf}$, which we parametrize in terms of the angles $\omega_{vf}$, encoding the relative embedding of the tetrahedra in a vertex, and the spinors $\ket{z_{ef}}$, one for each edge of each face, encoding their shape. Figure~\ref{fig:notation} summarizes this notation.
\subsection{Wedge holonomies and local flatness}
\label{sec:local-flatness}
\begin{figure}[H]
    \centering
    \includegraphics[scale=1]{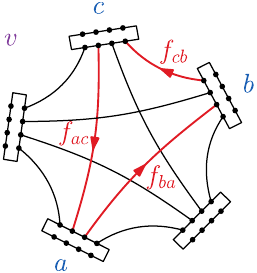}
    \caption{\label{fig:3-cycle} The 3-cycle in vertex $v$ of wedges $f_{ac}, f_{cb}, f_{ba}$ forming a closed loop around the vertex. The parallel transport around the loop is trivial.}
\end{figure}

As mentioned before, the wedge holonomies $g_{vf}$ outnumber the $4$ independent $\SL(2,\C)$ holonomies (one per edge, minus one overall gauge) needed to describe parallel transport inside a flat vertex. We remove this redundancy by requiring the parallel transport around every closed loop of wedges to be trivial, i.e.\ by requiring the two-complex to be built from flat building blocks. Since every loop in a vertex is generated by composing \emph{3-cycles} of three consecutive wedges (see Figure~\ref{fig:3-cycle}), it suffices to impose triviality on 3-cycles alone\footnote{Of the $10$ 3-cycles in a simplicial vertex, only $6$, the fundamental cycles, are independent. These reduce the number of independent variables from $10$ to $4$, as expected.},
\begin{equation}
    \label{eq:local-flatness}
    G_{vabc} \equiv g_{v f_{ac}}\, g_{v f_{cb}}\, g_{v f_{ba}} = \Id \, ,
\end{equation}
where $vf_{ba}$ is the wedge shared by edges $a$ and $b$. This is \emph{local flatness} as introduced in \cite{Dona:2022hgr}.
Local flatness \eqref{eq:local-flatness} strongly constrains the spinors of the three wedges involved (after a lengthy but straightforward computation, reported in \cite{Dona:2022hgr}) and fixes completely the complex angle of each wedge:
\begin{equation}
    \label{eq:omega-flatness}
    \omega_{vf_{ba}} = \varepsilon_v \theta_{vf_{ba}}^{c}(\{z_{vf}\}) - i\, \xi_{vf_{ba}}^{c}(\{z_{vf}\}) \, .
\end{equation}
The angle $\theta_{vf_{ba}}^c$ is the generalized dihedral angle between the tetrahedra at edges $a$ and $b$, and $\xi_{vf_{ba}}^c$ is the twist angle between their framed planes, both functions of the spinors alone. Explicitly
\begin{equation}
    \label{eq:boost-twist}
    \begin{aligned}
    \cosh\theta_{vf_{ba}}^{c} &= \frac{-|\brasket{\tilde z_{vf_{bc}}}{\tilde z_{vf_{ac}}}|^{2}+|\braket{\tilde z_{vf_{ab}}}{z_{vf_{bc}}}|^{2}|\braket{z_{vf_{ab}}}{z_{vf_{ac}}}|^{2}+|\brasket{z_{vf_{bc}}}{\tilde z_{vf_{ab}}}|^{2}|\brasket{z_{vf_{ab}}}{z_{vf_{ac}}}|^{2}}{2\left|\braket{z_{vf_{bc}}}{\tilde z_{vf_{ab}}}\,\brasket{z_{vf_{bc}}}{\tilde z_{vf_{ab}}}\,\braket{z_{vf_{ab}}}{z_{vf_{ac}}}\,\brasket{z_{vf_{ab}}}{z_{vf_{ac}}}\right|} \, , \\[0.75em]
    \xi_{vf_{ba}}^{c} &= \arg\left(\frac{\brasket{z_{vf_{bc}}}{\tilde z_{vf_{ab}}}\,\braket{\tilde z_{vf_{ab}}}{z_{vf_{bc}}}}{\braket{z_{vf_{ac}}}{z_{vf_{ab}}}\,\sbraket{z_{vf_{ab}}}{z_{vf_{ac}}}}\right) \, .
    \end{aligned}
\end{equation}
Geometrically, the scalar products of the spinors give the three-dimensional dihedral angles between the framed triangles they represent in a tetrahedron. The boost angle $\theta_{vf_{ba}}^c$ is then reconstructed from these angles by the spherical cosine law, using the triangle labelled by $c$ as reference. The twist $\xi_{vf_{ba}}^c$ is the residual rotation needed to align the two edge directions in the triangle after the parallel transport has aligned the two framed planes.

Furthermore, $\omega_{vf_{ba}}$, the complex angle of a single wedge holonomy, cannot depend on $c$. When we require consistency across all choices of $c$ we are forced to impose an angle-matching condition on the spinors, making $\theta_{vf_{ba}}$ and $\xi_{vf_{ba}}$ separately independent of $c$. Local flatness on all 3-cycles \emph{is} the shape-matching condition \cite{Dona:2020yao}. A similar result was observed in \cite{Anza:2014tea,Langvik:2016hxn}, in a different context, through an analogous calculation. It forces the triangle shapes seen from different tetrahedra to coincide and allows the tetrahedra to be embedded in four dimensions as a Lorentzian $4$-simplex. Here $\theta_{vf_{ba}}$ are positive dihedral boost angles, $\xi_{vf_{ba}}$ are the twists between the triangle frames, and $\varepsilon_v=\pm1$ records the orientation branch of the reconstructed $4$-simplex.

To be precise, not every solution of \eqref{eq:local-flatness} corresponds to a Lorentzian $4$-simplex geometry: a class of extra solutions, the so-called topological sector (traditionally known as vector geometries and Euclidean $4$-simplices), exists as well. We do not discuss them further and refer to \cite{Dona:2022hgr} for details, restricting from now on to the geometric Lorentzian sector.

In summary, local flatness of the wedge holonomies forces both the shape matching of the spinors' framed triangles, reconstructing a $4$-simplex geometry from the wedge data, and fixes the complex angles as $\omega_{vf}=\varepsilon_v\theta_{vf}-i\xi_{vf}$. Thus $\Re\omega_{vf}=\varepsilon_v\theta_{vf}$ and $\Im\omega_{vf}=-\xi_{vf}$.

Adding the local flatness constraints to the wedge description, as a set of delta functions, restricts the enlarged wedge-holonomy variables to the usual half-edge-holonomy data within each vertex. Exponentiating these delta functions with $\SL(2,\C)$-algebra-valued Lagrange multipliers $\mu_{vabc}$, the semiclassical constrained wedge action reads:
\begin{equation}
    \label{eq:actionEPRL}
    S^{\mathrm{EPRL-FK}} = i\sum_{f} j_{f} \sum_{v\subset f}\left( \gamma \, \Re\,\omega_{vf} + \Im\,\omega_{vf} \right) + i\sum_{vabc} \, \Tr\left( \mu_{vabc}\, G_{v abc} \right) \, .
\end{equation}
The multipliers $\mu_{vabc}$ are integrated over the algebra of $\SL(2,\C)$ with its standard Lebesgue measure invariant under conjugation, so that formally $\int \d\mu_{vabc}\, e^{i\Tr\left(\mu_{vabc}\, G_{vabc}\right)} \propto \delta(G_{vabc})$, reproducing the local flatness delta function at $G_{vabc} = g_{v f_{ac}}\, g_{v f_{cb}}\, g_{v f_{ba}} = \Id$. This Lie-algebra representation of the delta function is used here near the identity branch. The geometrically equivalent $G_{vabc}=-\Id$ branch of the double cover is not included in this local analysis as is customary in the derivation of the spinfoam amplitudes for BF theories.

This form has three practical advantages: it is linear in the spins and wedge angles, so the semiclassical analysis is straightforward; the local-flatness constraint depends only on the holonomies, not on the spins $j_f$; and variation with respect to the Lagrange multipliers $\mu_{vabc}$ returns the local-flatness condition \eqref{eq:local-flatness} on the chosen branch.

\subsection{Symmetries of the action}
\label{sec:symmetries}

A trivial redundancy comes from the parametrization itself: each wedge holonomy $g_{vf}$ is built from two unit spinors and a complex angle, eight real parameters for an element of the six-real-dimensional group $\SL(2,\C)$ (automatically unimodular, see Appendix~\ref{app:spinors}). Two of them are in fact redundant. There is a $\mathrm{U}(1)\times\mathrm{U}(1)$ freedom to shift the phases of $\ket{z_{vf}}, \ket{\tilde z_{vf}}$ against $\Im\,\omega_{vf}$ at each wedge independently, leaving $g_{vf}$ unchanged: explicitly, this phase action is
\begin{equation}
\label{eq:phase-act}
    (\ket{z_{vf}},\ket{\tilde{z}_{vf}},\operatorname{Im}\omega_{vf})\longmapsto(e^{i\phi}\ket{z_{vf}},e^{i\tilde{\phi}}\ket{\tilde{z}_{vf}},\operatorname{Im}\omega_{vf}-\phi-\tilde{\phi})\, , \qquad \phi,\tilde\phi\in\R/2\pi\mathbb{Z} \, .
\end{equation}
Physically, this means that we can always reabsorb a change of frame of the source and target triangles into a change of twist, without changing the geometry encoded in the wedge holonomy.

The action~\eqref{eq:actionEPRL} is also invariant under this parametrization change on the full two-complex, although the invariance is not manifest. Each spinor $\ket{z_{ef}}$ is shared by two consecutive wedges of the same face. Rephasing it shifts the imaginary angles of the two wedge holonomies equally and oppositely, $\delta\Im\omega_{vf}=-\delta\Im\omega_{v'f}$, leaving the face's total angle, and its contribution to the action, unchanged.

A genuine symmetry follows from the $\SU(2)$ covariance of the wedge holonomy \eqref{eq:wedge-param}: rotating the source spinor, $\ket{z_{vf}} \to u\ket{z_{vf}}$ (and correspondingly $\sket{z_{vf}}\to u\sket{z_{vf}}$), multiplies $g_{vf}$ by $u^\dagger$ \emph{on the right}; rotating the target spinor, $\ket{\tilde z_{vf}} \to u\ket{\tilde z_{vf}}$, multiplies $g_{vf}$ by $u$ \emph{on the left}. For independent $u,u'\in\SU(2)$ acting on the target and source spinor respectively,
\begin{equation}
    \label{eq:su2-frame-gauge}
    g_{vf} \; \longmapsto \; u\, g_{vf}\, u'^{\dagger} \, .
\end{equation}
Physically, $u$ and $u'$ rotate the reference frame within the target and source framed plane, respectively.

Now fix a single edge, say edge $b$ of the 3-cycle \eqref{eq:local-flatness}, and rotate \emph{every} spinor attached to it by the same $u_b \in \SU(2)$. Of the three wedges of the cycle, only the two that meet at $b$ are affected: $b$ is the target of $g_{vf_{ba}}$ and the source of $g_{vf_{cb}}$, so by \eqref{eq:su2-frame-gauge}
\begin{equation}
    \label{eq:edge-gauge-cycle}
    g_{vf_{ba}} \; \longmapsto \; u_b\, g_{vf_{ba}} \, , \qquad g_{vf_{cb}} \; \longmapsto \; g_{vf_{cb}}\, u_b^\dagger \, ,
\end{equation}
while $g_{vf_{ac}}$, which does not touch edge $b$, is left unchanged. The Lagrange-multiplier term of the action \eqref{eq:actionEPRL} then transforms as
\begin{equation}
    \label{eq:cycle-invariance}
    \Tr \left(\mu_{vabc}\, g_{vf_{ac}}\, g_{vf_{cb}}\, g_{vf_{ba}}\right) \; \longmapsto \;  \Tr\left(\mu_{vabc}\, g_{vf_{ac}}\, g_{vf_{cb}}\, u_b^\dagger u_b\, g_{vf_{ba}}\right) = \Tr\left(\mu_{vabc}\, g_{vf_{ac}}\, g_{vf_{cb}}\, g_{vf_{ba}}\right) \, ,
\end{equation}
the $u_b^\dagger u_b$ produced at the shared edge $b$ cancels immediately, leaving both the bare product \eqref{eq:local-flatness} and this term of the action invariant, without ever needing to touch $\mu_{vabc}$.

Edge $a$ is different: it sits at the two \emph{ends} of the cyclic product, rather than between two adjacent factors, so the transformation does not cancel on the spot but survives as an overall conjugation,
\begin{equation}
    \Tr \left( \mu_{vabc} g_{vf_{ac}}\, g_{vf_{cb}}\, g_{vf_{ba}} \right)\; \longmapsto \; \Tr \left( \mu_{vabc}u_a\, g_{vf_{ac}}\, g_{vf_{cb}}\, g_{vf_{ba}}\, u_a^\dagger \right) \, .
\end{equation}
Invariance of the action still holds, but now needs cyclicity of the trace together with the change of variables $\mu_{vabc} \to u_a^\dagger\, \mu_{vabc}\, u_a$, which leaves the conjugation-invariant measure over $\mu_{vabc}$ unchanged.

So the Lagrange-multiplier term of the action is invariant under an $\SU(2)$ rotation $u_e$ of the spinors at edge $e$. The first term of the action, meanwhile, is unaffected: it depends only on $\omega_{vf}$, which does not transform under this spinor rotation. The same argument applies to every edge and every 3-cycle through it, so the full action \eqref{eq:actionEPRL} is invariant under an independent $\SU(2)$ rotation $u_e$ at each edge $e$ of the vertex. Explicitly, writing $\ket{z_{ef_1}},\dots,\ket{z_{ef_4}}$ for the four normalized spinors identified at edge $e$, this $SU(2)$ action is
\begin{equation}
\label{eq:su2-act}
    (\ket{z_{ef_1}},\ket{z_{ef_2}},\ket{z_{ef_3}},\ket{z_{ef_4}})\longmapsto(u_e\ket{z_{ef_1}},u_e\ket{z_{ef_2}},u_e\ket{z_{ef_3}},u_e\ket{z_{ef_4}}) \, , \qquad u_e\in\SU(2) \, .
\end{equation}

This edge $\SU(2)$ symmetry is the same gauge freedom used in the EPRL-FK construction to relate the spinfoam boundary data to the kinematical states of Loop Quantum Gravity.

\subsection{The equations of motion of the EPRL-FK model and the flatness problem}
\label{sec:eom}

In the semiclassical path-integral analysis, classical equations of motion follow from extremizing the action that appears in the exponent of the amplitude. Here we pin down exactly why the standard semiclassical action \eqref{eq:actionEPRL} produces the flatness equation instead of the Regge equations, before any continuum limit is taken.

Varying the EPRL-FK action in wedge variables, \eqref{eq:actionEPRL}, with respect to the Lagrange multipliers $\mu_{vabc}$ returns the local flatness constraint \eqref{eq:local-flatness}. We already discussed it in Section~\ref{sec:local-flatness}. Varying instead with respect to the wedge holonomy parameters \eqref{eq:wedge-param}, namely the complex angles $\omega_{vf}$ and the spinors shared between consecutive wedges, fixes the multipliers themselves in terms of the spins $j_f$ and the geometry encoded in the spinors.

In terms of the wedge multiplier $\mu_{vf} \equiv \sum_{vabc\supset vf}\mu_{vabc}$ the solution has a simpler geometric interpretation. On shell of local flatness, $\mu_{vf}$ parallel transports around its face like the source-triangle normal algebra element (see Appendix~\ref{app:spinors})
\begin{equation}
    \label{eq:normal-algebra}
    N_{vf} = \vec n_{vf}\cdot\vec\sigma = \sket{z_{vf}}\sbra{z_{vf}}-\ket{z_{vf}}\bra{z_{vf}} \, .
\end{equation}
Since $\vec n_{vf}$ depends only on the identified spinor $z_{vf}=z_{ef}$, we likewise drop the vertex label and write $N_{ef}$ from now on,
\begin{equation}
    \label{eq:mu-fixed}
    \mu_{vf} = \frac{1}{2} j_{f} \left( \gamma - i \right) N_{ef} \, .
\end{equation}
The wedge multiplier is $-i$ times the $\gamma$-simple bivector of the triangle\footnote{$\vec{\Pi}=\vec{E} + i \vec{B}$ with $\vec B=\gamma\vec E$, where $\vec{E}$ and $\vec{B}$ are the electric and magnetic parts of the bivector.} (see \cite{Dona:2020xzv} for a detailed discussion).

\begin{figure}[H]
\centering
\includegraphics[scale=1]{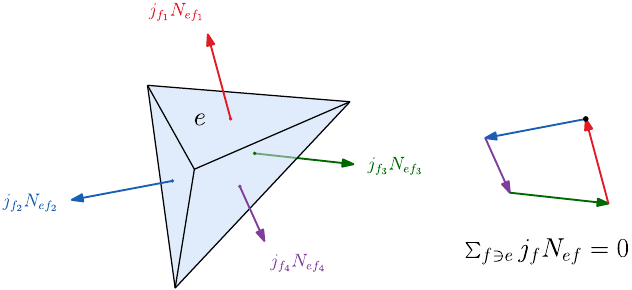}
\caption{\label{fig:closure} The tetrahedron at the edge $e$, with the outward normal of each of its four triangles weighted by the corresponding spin, $j_fN_{ef}$ (left). Translated tip-to-tail, the same four vectors close into a loop (right): this is the geometric closure condition \eqref{eq:closure-spinorial}.}
\end{figure}

The wedge multipliers $\mu_{vf}$ are not independent. At each edge $e$ of the vertex $v$ (taken as the source edge of every wedge below), their equations imply a geometric closure relation (see Appendix~\ref{app:eom-details}). Substituting \eqref{eq:mu-fixed} and dropping the common nonzero factor $(\gamma - i)$ gives
\begin{equation}
    \label{eq:closure-spinorial}
    \sum_{f \ni e} \mu_{vf} = 0
    \qquad\longleftrightarrow\qquad
    C^{\mathrm{geo}}_e \equiv \sum_{f \ni e} j_{f}\, N_{ef} = 0 \, .
\end{equation}
This is the geometric closure condition for the full bivectors of the tetrahedron at edge $e$ \cite{Livine:2011gp}, illustrated in Figure~\ref{fig:closure}. It should not be confused with the quantum Gauss closure imposed on EPRL-FK intertwiners, nor with closure of coherent-intertwiner labels. The point is that $N_{ef}$ is extracted from the wedge-holonomy geometry, so $C^{\mathrm{geo}}_e=0$ ties the spin $j_f$ to that geometry off shell. 

Finally, varying the action with respect to the spins $j_f$ returns the flatness equation after dividing out the common nonzero factor of $i$,
\begin{equation}
    \sum_{v\subset f}\left( \gamma \, \Re\,\omega_{vf} + \Im\,\omega_{vf} \right) = 0  \, .
\end{equation}
In this stationary-phase analysis we treat the spins as continuous variables. This is equivalent to the discrete-spin treatment under mild assumptions on the oscillation frequency \cite{Dona:2025snr}. Equivalently, one may keep the spin sum and use Poisson resummation. The same large-spin equations are obtained provided the periodicity of the variables conjugate to the spins is retained.

\begin{figure}[H]
\centering
\includegraphics[scale=1.0]{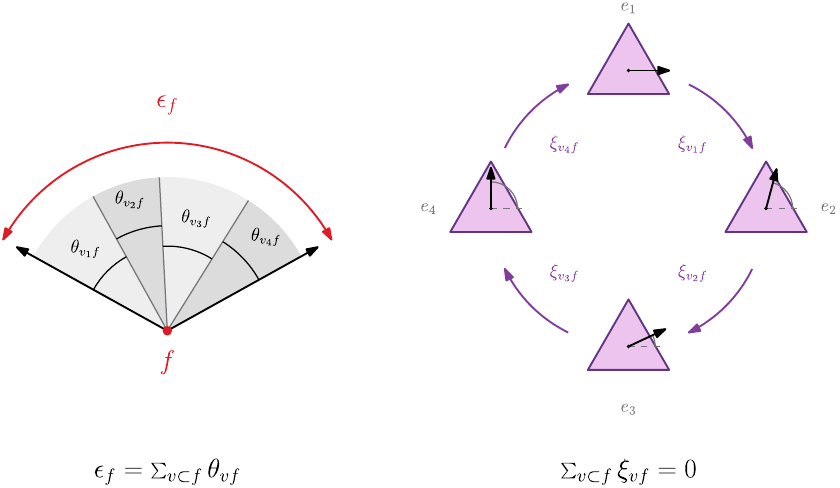}
\caption{\label{fig:deficit-angle} The two angles characterizing the holonomy around a face $f$. Left: seen in the plane orthogonal to the hinge, the wedges forming the face dual to the hinge fan out with boost angles $\theta_{vf}$, and the frame arrives boosted with respect to the one it started from by the deficit angle $\epsilon_f$. Right: seen in the plane of $f$, the framing of the triangle (black arrow, against a fixed reference in gray) is twisted by $\xi_{vf}$ at each wedge, and returns to itself after the full loop, so the twists cancel. The angles are exaggerated for readability.}
\end{figure}

On shell of local flatness, the equation reduces to the vanishing of the deficit angle around the face. On the globally consistent orientation branch,
\begin{equation}
\sum_{v\subset f} \left( \gamma \, \Re\,\omega_{vf} + \Im\,\omega_{vf} \right)
= \gamma\sum_{v\subset f}\varepsilon_v\theta_{vf}
- \sum_{v\subset f}\xi_{vf}
= \gamma \epsilon_f = 0 \, .
    \label{eq:deficit-angle-definition}
\end{equation}
Here we have defined the deficit angle as
\begin{equation}
    \epsilon_f \equiv \sum_{v\subset f}\varepsilon_v\theta_{vf}\, ,
\end{equation}
and used the fact that the sum of twist angles around an internal face vanishes identically, $\sum_{v\subset f}\xi_{vf}=0$ (see \cite{Dona:2020xzv} for a detailed discussion). This is illustrated in Figure~\ref{fig:deficit-angle}.

\subsection{What's the problem with EPRL-FK equations of motion?}

Local flatness already fixes the shape data of the $4$-simplex at each vertex. From the scalar products of the spinors we can build, for each tetrahedron, the Gram matrix of its four face normals. The area ratios needed to close the tetrahedron are then fixed by this Gram matrix, up to a single overall scale. However, local flatness alone does not tie the independent spins to the area ratios selected by this shape. In area-angle Regge calculus \cite{Dittrich:2008va} this link is provided by closure, entering the action through a Lagrange multiplier coupled directly to the areas, so that varying with respect to them gives a non-trivial equation intertwined with closure. In its absence, the reconstructed shape and the area variables variations remain decoupled.

In the constrained-wedge description above, geometric closure holds for geometric critical configurations but does not enter the spin variation through an independent multiplier. The spin variables therefore remain independent from the shape ones, which produces the flatness equation rather than the Regge equations. The same happens in area-angle Regge calculus if shape matching is imposed but closure is not: the action is linear in the areas, and varying them again returns only the flatness equation, with no coupling to the shape fixed by the angles.

The space of areas and angles is larger than the space of length-Regge geometries. Shape matching and geometric closure are the two constraints that single out the length-Regge subspace. If both constraints are imposed strongly on the unconstrained area-angle action, through Lagrange multipliers coupled respectively to the angles and to the areas, the resulting equations of motion reproduce length Regge calculus in the nondegenerate sector. This motivates imposing geometric closure as an independent constraint in the wedge variables.

\section{Dynamical versus strong constraints}
\label{sec:dynamical-vs-strong}
Before returning to spinfoams, let us recall a general point about constrained variational principles. A condition that follows from the equations of motion is not equivalent, in general, to the same condition imposed from the start with a Lagrange multiplier. 

Recall the general logic of constrained extremization. Given a function $f(q)$ that one wishes to extremize subject to a condition $g(q)=0$, the Lagrange multiplier method introduces an auxiliary variable $\lambda$ and instead extremizes the unconstrained function $L(q,\lambda) = f(q) + \lambda\, g(q)$ freely over both $q$ and $\lambda$. Stationarity with respect to $\lambda$ simply returns the constraint $g(q)=0$, while stationarity with respect to $q$ gives $\nabla f(q) + \lambda\, \nabla g(q) = 0$. Crucially, $\lambda$ is not fixed in advance: its value is determined together with $q$ by solving these two conditions simultaneously.

\begin{figure}[H]
    \centering
    \includegraphics[width=0.6\linewidth]{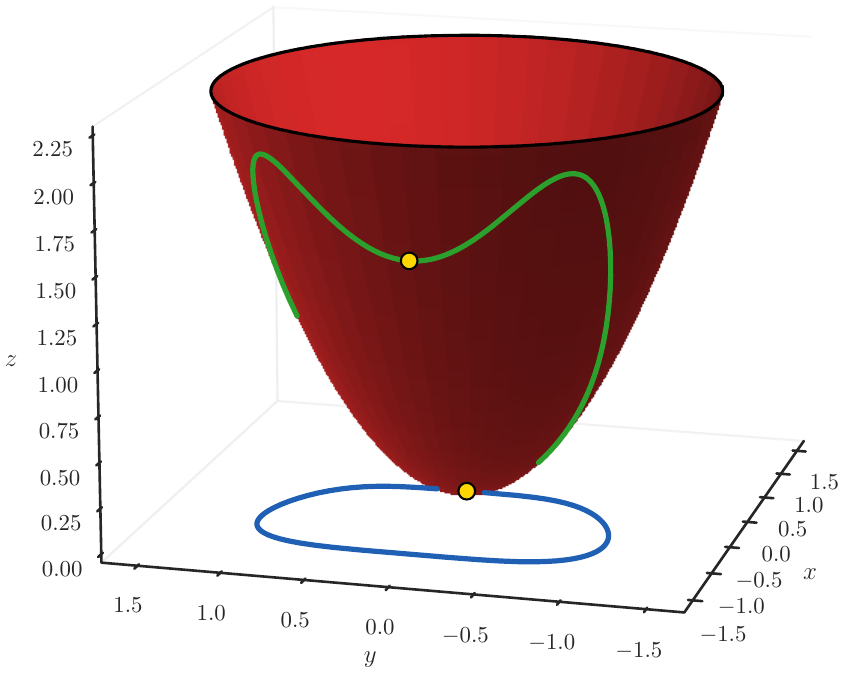}
    \caption{\label{fig:3d-surface} Illustration of the solution space for the unconstrained and constrained systems. The red surface represents the action $x^2+y^2$ of the unconstrained system. Its minimum is located at the origin, ans is highlighted in yellow. The blue line represents the constraint $x^4+y^4+2x=0$. When lifted onto the surface, it becomes the green line, which now has one additional minimum.}
\end{figure}

As a simple illustration, consider the problem of finding the global minimum of the function $x^2 + y^2$ subject to the constraint $x^4 + y^4 + 2x = 0$. The unconstrained minimum is clearly at the origin $(0,0)$. Imposing the constraint introduces additional solutions that satisfy both the original equations of motion and the constraint, demonstrating that the constrained system can have more solutions than the unconstrained one. See Figure~\ref{fig:3d-surface} for a visual representation.

\begin{samepage}
Analogously, for a dynamical system with dynamical variables $q$ and unconstrained action $S_0[q]$, whose equations of motion we denote $E_i[q] \equiv \delta S_0[q]/\delta q^i$. Suppose a condition $F[q]=0$ is not itself an independent constraint but a linear combination of these equations of motion,
\begin{equation*}
   F[q] = \alpha^i[q] E_i[q]=0 \, .
\end{equation*}
\end{samepage}

Promoting $F$ to an independently enforced constraint means introducing a real Lagrange multiplier $\lambda(t)$ and defining the extended action
\begin{equation*}
    S_{\mathrm{ext}}[q,\lambda] = S_0[q] + \int \d t\, \lambda\, F[q] \, ,
\end{equation*}
need not recover the original theory. Varying the extended action with respect to $q^i$ gives $E_i[q] + \lambda\, \delta F[q]/\delta q^i = 0$, so its equations of motion may admit solutions with
\begin{equation*}
    E_i[q] = -\lambda\, \frac{\delta F[q]}{\delta q^i} \neq 0 \, ,
\end{equation*}
violating the equations of motion of $S_0$ alone, precisely because $F=0$ does not fix the multiplier's own dynamics. The original solutions can be embedded in the modified problem by choosing an appropriate multiplier (e.g. $\lambda=0$ recovers them exactly), but the modified variational problem may contain additional solutions. Appendix~\ref{app:particle-angular-momentum} illustrate strict enlargement in a simple mechanical example, a free particle whose conservation of angular momentum is exactly such an on-shell consequence.

The constrained-wedge description is in this situation. Geometric closure,
\begin{equation}
    C^{\mathrm{geo}}_e = 0 \, ,
\end{equation}
holds for the geometric critical configurations of \eqref{eq:actionEPRL}, but it is not an independently enforced constraint in that action. The absence of independently imposed holonomy-flux compatibility leaves the area variables independent in the spin variation and therefore produces the flatness equation in this constrained-wedge description. We now modify the constrained semiclassical problem by adding geometric closure to the action with its own multiplier $\lambda_e$, on the same footing as local flatness. As in the general discussion above, this can enlarge the space of solutions. The question is whether the enlarged variational problem provides precisely Regge geometries.
\begin{figure}[H]
    \centering
    \includegraphics[width=0.3\linewidth]{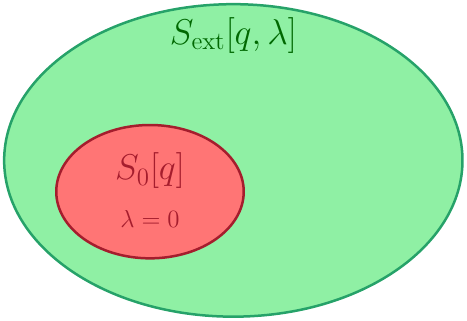}
    \caption{\label{fig:solution-space} Solution spaces of the unconstrained and constrained systems. The red region is the set of solutions $S_0[q]$ of the unconstrained action. The green region is the set of solutions of the extended action $S_{\mathrm{ext}}[q,\lambda]$, projected onto $q$. Setting $\lambda=0$ embeds the former inside the latter, but the extended variational problem generically admits further solutions with $\lambda\neq0$ lying outside this embedded copy.}
\end{figure}

\section{Imposing geometric closure strongly}
\label{sec:closure-spinors}
We define the constrained action by introducing, for each edge $e$, a Lagrange multiplier for the geometric-closure condition \eqref{eq:closure-spinorial}. Since $C^{\mathrm{geo}}_e=\sum_{f\in e} j_f N_{ef}$ is Hermitian and traceless (the algebra elements are defined in \eqref{eq:normal-algebra}), geometric closure is a genuinely $3$-real-dimensional constraint, so the multiplier $\lambda_{e}$ is taken $\SU(2)$ algebra-valued. The resulting action reads
\begin{equation}
    \label{eq:actionEPRL-closure}
    S^{\EPRLclosed} = i\sum_{f} j_{f} \sum_{v\subset f}\left( \gamma \, \Re\,\omega_{vf} + \Im\,\omega_{vf} \right)
    + i\sum_{vabc} \, \Tr\left( \mu_{vabc}\, G_{vabc} \right) + i\sum_{e} \, \Tr\left( \lambda_{e}\, C^{\mathrm{geo}}_e \right) \, .
\end{equation}

Because of the Lagrange multipliers $\lambda_{e}$, variation with respect to the spins $j_f$ no longer imposes flatness of the faces
\begin{equation}
    -i\frac{\delta S^{\EPRLclosed}}{\delta j_f}
    =
    \gamma \epsilon_f + \sum_{e\subset f} \Tr\left( \lambda_{e}\, N_{ef} \right)
    =0 
    \, , \quad
    \centernot\implies \quad \epsilon_f = 0
    \, .
\end{equation}
This is analogous to what happens in area-angle Regge calculus \cite{Dittrich:2008va}: when the areas are varied independently, the deficit angles are not set to zero but are balanced by the Lagrange multipliers enforcing closure.

The main result of this section is that, on every regular, nondegenerate, consistently oriented Lorentzian geometric component with spacelike tetrahedra, the gauge classes of saddle points of $S^{\EPRLclosed}$ are in one-to-one correspondence with solutions of the length-Regge equations for the internal edge variables, with the boundary data held fixed. The proof proceeds in three steps:
\begin{enumerate}
    \item We show that the equations obtained by varying the ambient variables and the Lagrange multipliers are equivalent to stationarity of $S^{\EPRLwedge}$ \eqref{eq:EPRL-wedge} under variations tangent to the constraint surface.
    \item We restrict to the geometric sector and explicitly reconstruct the squared edge lengths $s_p$ from the constrained data $(j_f,z_{ef},\omega_{vf})$. We then show that this reconstruction is gauge invariant and construct its inverse, thereby proving that $\mathcal C_{\mathrm{geo}}/\mathcal G$ is diffeomorphic to the space $\mathcal{U}_{\mathrm{Regge}}$ of Lorentzian Regge geometries on the chosen branch.
    \item We use the squared edge lengths $s_p$ as coordinates on this quotient and show that stationarity along the corresponding tangent variations $D_p$ is equivalent to the length-Regge equations.
\end{enumerate}
Thus the geometric-closure constraint replaces the independent spin variations by edge lengths variations: instead of imposing vanishing deficit angles, the variational principle produces the Regge equations of motion.

\subsection{The constrained variational problem}
\label{sec:constrained-variational-problem}

The method of Lagrange multipliers ensures that the variation of a function restricted to a submanifold defined by sufficiently regular constraints is equivalent to the variation of the corresponding \emph{Lagrangian function}, obtained by augmenting the original function with the constraints coupled to Lagrange multipliers. From a physical perspective, this means that we can construct an equivalent theory by extending the original configuration space and defining an action in which the constraints are coupled to Lagrange multipliers. The two theories are equivalent in the sense that they have the same physical trajectories, i.e., the saddle points of the respective actions, while the multipliers themselves carry no independent physical content. Appendix~\ref{app:particle-circle} illustrates this criterion in a simple mechanical example: a particle constrained to move on a circle.

\medskip

We first formulate the constrained variational problem as a local proposition. On a regular component of the constraint surface, the stationary points of the geometrically closed EPRL-FK action are precisely the stationary points of $S^{\EPRLwedge}$ restricted to that surface. We then quotient out the gauge directions and determine the saddle points modulo gauge transformations.

\paragraph{Ambient variables and constraints.}
We collect the spins, the normalized spinors, and the complex wedge angles in
\begin{equation}
    q=\left\{j_f,z_{ef},\omega_{vf}\right\} \, ,
\end{equation}
and denote by $\mathcal{Q}$ the space of all such data. Notice that $\mathcal{Q}$ is finite dimensional. The wedge holonomies $g_{vf}=g_{vf}(z,\omega)$ depend on them through \eqref{eq:wedge-param}, and the part of the action carrying no Lagrange multiplier is the wedge action itself, $S^{\EPRLwedge}$, \eqref{eq:EPRL-wedge}. As above, our stationary-point analysis treats the spins as continuous large-spin variables. In what follows, we ignore boundary terms and work only with internal variations. For a simplicial complex with boundary, the boundary data are held fixed.

\smallskip

For every edge $e$, geometric closure gives the constraint $C^{\mathrm{geo}}_e(q)=0$. Local flatness is imposed independently at each vertex $v$. We choose a maximal tree in the graph of wedge holonomies. Its six chords define six independent fundamental cycles $abc$, on which $G_{vabc}(q)=\Id$. The surface defined by the constraints is
\begin{equation}
    \mathcal C
    \equiv
    \left\{
        q \in \mathcal{Q}\;\middle|\;
        C^{\mathrm{geo}}_e(q)=0\, ,\quad
        G_{vabc}(q)-\Id = 0
    \right\}\, .
    \label{eq:constraint-surface}
\end{equation}
We call the collection of the constraints $C$ seen as a map $C:\mathcal{Q}\to \mathbb{R}^m$, where $m$ depends on the specific 2-complex. Notice that in any case $\mathrm{dim}(\mathcal{Q})=n>m$. Thus, $\mathcal{C}=C^{-1}(0)$. The constraints may be redundant, but this does not affect the analysis that follows. Moreover, $\mathcal C$ is invariant under the gauge symmetry transformations we described in Section~\ref{sec:symmetries}.

The extended action is $S^{\EPRLclosed}$, \eqref{eq:actionEPRL-closure}. From now on, we suppose that $C$ is a map with constant rank in an open neighborhood of any point $q\in \mathcal{C}$. Otherwise, we restrict ourselves to the points $q\in\mathcal{C}$ admitting an open neighborhood $U_q\subset \mathcal{Q}$ such that $\mathrm{rk}(d_{q'}C)=r$ for every $q'\in U_q$ and some $r\leq m$ \footnote{Here $d_qC$ is the tangent map of $C$ at $q$:
\begin{equation*}
    d_qC:T_q\mathcal{Q}\to \mathbb{R}^m\, .
\end{equation*}}. The rank theorem ensures that the collection of such points (if not empty) is a submanifold with dimension $n-r$. We continue to denote this submanifold by $\mathcal{C}$. Later, we will quotient out the gauge directions from the geometric part of $\mathcal{C}$ when comparing it with the Regge configuration space.

\paragraph{Constrained variations and saddle points.}
Since $\mathcal{Q}$ is a finite-dimensional smooth manifold, the variation of a generic smooth function $f:\mathcal{Q}\to \R$ is the standard differential $df$, and $q$ is a saddle point if the differential at $q$ is zero:
\begin{equation}
    d_qf=0 \, .
\end{equation}
In case we are looking for a saddle point of $f$ constrained on the submanifold $\mathcal{C}$, we are looking for
\begin{equation}
    \d_q(f|_\mathcal{C})=0 \, ,
\end{equation}
where $\d$ is the differential on $\mathcal{C}$, distinct from the ambient differential $d$ on $\mathcal{Q}$ used above. The latter condition is equivalent to
\begin{equation*}
    \mathrm{ker}(d_qf)\supset T_q\mathcal{C} \, .
\end{equation*}
By hypothesis and by the rank theorem,
\begin{equation*}
    T_q\mathcal{C}=\ker(d_qC) \, .
\end{equation*}
hence
\begin{equation*}
    \mathrm{ker}(d_qf)\supset \ker(d_qC) \, .
\end{equation*}
As a simple exercise in linear algebra, this condition holds if and only if $d_qf$ is a linear combination of the components of the constraint $C$, namely, there exist some $\Lambda_i\in \mathbb{R}$ with $i\in\{1,\dots,m\}$ such that
\begin{equation}
\label{eq:lagr-mult}
    d_qf=\sum_{i=1}^m\Lambda_i\,d_qC_i \, .
\end{equation}
Finally, the pair of conditions for $q$ to be a constrained saddle point, given by \eqref{eq:lagr-mult} and $C(q)=0$, is equivalent to the variation with respect to $q$ and $\Lambda$ of
\begin{equation}
    \mathcal{L}(q,\Lambda)=f(q)+\Lambda\cdot C(q) \, .
\end{equation}
where $\cdot$ is the dot product in $\R^m$. Notice that the exact value of the Lagrange multipliers does not play any role: \eqref{eq:lagr-mult} may be rewritten without using them as
\begin{equation*}
    d_qf\wedge d_qC_1\wedge\dots\wedge d_qC_m=0 \, .
\end{equation*}

\smallskip

We have thus established, by a straightforward application of the method of Lagrange multipliers, that on every regular component of the constraint surface,
\begin{equation}
    q \text{ is a stationary point of }S^{\EPRLclosed}
        \quad \Leftrightarrow \quad
        q\text{ is a stationary point of }
        S^{\EPRLwedge}\big|_{\mathcal C} \, .
    \label{eq:constrained-equivalence}
\end{equation}
This establishes the equivalence between the ambient constrained variational problem and the variational problem intrinsic to $\mathcal C$. The remaining task is to show that, in the nondegenerate Lorentzian sector and modulo gauge transformations, $\mathcal C$ is the space of Regge geometries. We must then identify its physical tangent directions and evaluate the equation of motion along them.

\paragraph{Gauge redundancy and orbit space.}
As shown in Section~\ref{sec:symmetries}, the action $S^{\EPRLwedge}\big|_{\mathcal C}$ has two independent families of gauge symmetries: the phase action \eqref{eq:phase-act}, an independent shift of phases on each wedge $(v,f)$ encoded by an action of $\U(1)\times \U(1)$, and the $\SU(2)$ action \eqref{eq:su2-act}, a single rotation $u_e\in\SU(2)$ acting simultaneously on the four normalized spinors associated with each dual edge $e$. These two may be collected in the smooth action of a single compact Lie group $\mathcal{G}$ defined as:
\begin{equation}
    \mathcal{G}=\prod_e \SU(2)\times\prod_{vf}(\U(1)\times \U(1)) \,.
\end{equation}
Therefore, we may formulate the variational principle directly on the orbit space $\mathcal{C}/\mathcal{G}$. Fortunately, $\mathcal{C}/\mathcal{G}$ is a smooth manifold, and $S^{\EPRLwedge}\big|_{\mathcal C}$ descends to a smooth function on the quotient. This follows from the fact that the action of $\mathcal{G}$ on $\mathcal{C}$ is proper and free. The action is proper because every continuous action of a compact group is proper. The action \eqref{eq:phase-act} is free because an element in the stabilizer of a point must, in particular, satisfy
\begin{equation*}
    \ket{z_{ef}}=e^{i\phi}\ket{z_{ef}} \,.
\end{equation*}
Since the spinor has unit norm, this equation is satisfied only if $e^{i\phi}=1$, and hence the stabilizer is trivial. Similarly, for the action \eqref{eq:su2-act}, an element $u_e\in\SU(2)$ in the stabilizer of a point must satisfy
\begin{equation*}
    u_e\ket{z_{ef_i}}=\ket{z_{ef_i}}\,,
\end{equation*}
which means that $u_e$ has eigenvalue $1$. Since the two eigenvalues of an element of $\SU(2)$ are complex conjugate phases, the other eigenvalue must also be $1$. Therefore, $u_e$ is the identity\footnote{Strictly speaking, one must also exclude mixed stabilizers, in which an $\SU(2)$ rotation is compensated by the phase action. In the nondegenerate sector considered here, such mixed stabilizers are trivial as well.}.

\smallskip

More generally, if $f$ is a $\mathcal{G}$-invariant function on $\mathcal{C}$, its stationary points are unions of gauge orbits, which become points (not necessarily isolated) in $\mathcal{C}/\mathcal{G}$. Indeed, let
\begin{equation*}
    f:\mathcal{C}\to\mathbb{R} 
\end{equation*}
be a $\mathcal{G}$-invariant function, i.e.,
\begin{equation}
f(g\triangleright q)=f(q)\, ,\quad \forall g\in\mathcal{G}\, ,\, q\in\mathcal{C}\, .
\end{equation}
Then $f$ descends to the quotient: there exists a unique function
\begin{equation*}
    \bar{f}:\mathcal{C}/\mathcal{G}\to\mathbb{R} \, ,
\end{equation*}
such that
\begin{equation}
\bar{f}([q])=f(q) \, .
\end{equation}
Equivalently,
\begin{equation*}
    \bar{f}\circ\pi=f\, ,
\end{equation*}
where $\pi:\mathcal{C}\to\mathcal{C}/\mathcal{G}$ is the quotient map. Differentiating this relation gives
\begin{equation}
\bar{\d}_{[q]}\bar{f}\circ \d_q\pi=\d_qf \, ,
\end{equation}
where $\bar{\d}$ denotes the differential on $\mathcal{C}/\mathcal{G}$. Since $\pi$ is a submersion, $\d_q\pi$ is surjective, and therefore
\begin{equation*}
\bar{\d}_{[q]}\bar{f}=0
\quad\Longleftrightarrow\quad
\d_qf=0 \, .
\end{equation*}
Consequently,
\begin{equation*}
q\text{ is a stationary point of }S^{\EPRLwedge}\big|_{\mathcal C}
\quad\Longleftrightarrow\quad
[q]\text{ is a stationary point of }\bar{S}^{\EPRLwedge}
\end{equation*}
where $\bar{S}^{\EPRLwedge}$ denotes the function induced by $S^{\EPRLwedge}\big|_{\mathcal C}$ on the orbit space $\mathcal{C}/\mathcal{G}$.

\subsection{Geometric identification of the constraint surface}
\label{sec:geometric-constraint-surface}
In the nondegenerate Lorentzian sector with spacelike tetrahedra, $\mathcal{C}$ modulo the gauge symmetries is the space of Regge geometries. To prove this, we reconstruct one scalar $s_p$ for every triangulation edge $p$ dual to the simplicial complex. Geometrically, $s_p$ is the squared length of $p$. These reconstructed quantities will only be used as local coordinates on the orbit space: we do not replace the spins, spinors, or wedge angles in the action with length variables. We work in spin-area units, so the triangle areas are represented directly by the spins $j_f$.

\paragraph{Restriction to the geometric sector.}
The complete constraint surface $\mathcal C$ also contains degenerate and nongeometric solutions \cite{Dona:2022hgr}. We therefore restrict to an open geometric sector
\begin{equation}
    \mathcal C_{\mathrm{geo}}\subset\mathcal C\, ,
\end{equation}
in which geometric closure reconstructs nondegenerate Euclidean tetrahedra and local flatness reconstructs nondegenerate Lorentzian four-simplices with spacelike boundary tetrahedra.
This sector is characterized by the inequality \begin{equation}\label{eq:ineq}
    |\cosh{\theta_{vf}}|>1 \, ,
\end{equation}
where the angle $\theta_{vf}$ is the dihedral boost angle as defined in \eqref{eq:boost-twist}.
The tetrahedra and their triangular faces are therefore spacelike and nondegenerate. We fix a parity branch and a globally consistent orientation branch. Thus the signs $\varepsilon_v$ appearing in Eq.~\ref{eq:phase-solution} are fixed choices, not additional continuous variables and not gauge directions. For every triangulation edge $p$, we assume that the incidence graph of the four-simplices containing $p$, with adjacency defined by shared tetrahedra containing $p$, is connected\footnote{This mild technical assumption ensures that the reconstructed edge geometry can be propagated consistently among all four-simplices incident on the edge. It is automatic when the simplicial complex is a triangulation of a genuine four-manifold (possibly with boundary).} Since it is not possible to pass continuously from a given orientation to another, they lie on different connected components of solutions of Eq.~\eqref{eq:ineq}. Therefore, being a geometric non-degenerate configuration as in Eq.~\eqref{eq:ineq} is an open condition in $\mathcal{C}$. Consequently, $\mathcal{C}_{\mathrm{geo}}$ is an open submanifold of $\mathcal{C}$, and
\begin{equation}
    T_q\mathcal C_{\mathrm{geo}}=T_q\mathcal C=\ker d_q C\, .
\end{equation}
Note that this restriction is a simplifying assumption about which part of the constraint surface we work on. The same assumption was used in the Riemannian case in \cite{Dittrich:2008va} to show that the area-angle Regge action reproduces the length-Regge equations of motion.

The gauge group $\mathcal G$ used below is the product of the continuous redundancies of the wedge parametrization. It contains independent edge-frame rotations $u_e\in\SU(2)$ acting on the four spinors at each dual edge, and spinor-phase redundancies compensated by shifts of $\Im\omega_{vf}$ so that the holonomies $g_{vf}$ are unchanged. The different parity and globally consistent orientation branches are not mapped into each other by the action of $\mathcal G$.

We now want to map points in $\mathcal{C}_{\mathrm{geo}}$ to points in the space $\mathcal{U}_{\mathrm{Regge}}$ of Lorentzian Regge geometries in which every four-simplex has spacelike boundary tetrahedra. We parametrize this space by a collection of positive numbers $s_p$, one for each dual edge of the complex, representing squared edge lengths and therefore subject to suitable inequalities. In particular, we want to reconstruct the edge lengths from $j_f$ and $\ket{z_{ef}}$ in such a way that $j_f$ and $N_{ef}$ can be interpreted, respectively, as the areas and normals of the reconstructed geometry. Since the orientation is fixed, $\omega_{ef}$ is fully determined by the spinors.

\paragraph{Reconstruction of the edge scalars.}
For every dual edge $e$, geometric closure $C^{\mathrm{geo}}_e=0$ and the three-dimensional Minkowski reconstruction theorem imply that the data $\{j_f,N_{ef}\}_{f\in e}$, with $N_{ef}=\vec n_{ef}\cdot\vec\sigma$, determine a unique Euclidean tetrahedron modulo the frame gauge at $e$. In $\mathcal C_{\mathrm{geo}}$, this tetrahedron is nondegenerate by definition. It has triangle areas $j_f$ and outgoing normals $\vec n_{ef}$, and therefore determines its six edge lengths.

\medskip

As mentioned in Section~\ref{sec:local-flatness}, local flatness $G_{vabc}=\Id$ is exactly the shape-matching condition on the chosen branch: two tetrahedra $e,e'$ of the same four-simplex $v$ sharing a triangle $f$ assign to it the same intrinsic angles and, since they also share the area $j_f$, the same side lengths. Geometric closure and local flatness glue the tetrahedra consistently along $f$. The five tetrahedra at $v$ then form the boundary of a nondegenerate Lorentzian four-simplex, with
\begin{equation}\label{eq:phase-solution}
    \omega_{vf}=\varepsilon_v\theta_{vf}-i\xi_{vf}\, ,
\end{equation}
where $\theta_{vf}$ is its positive dihedral boost angle, $\xi_{vf}$ is the twist relating the frames of the shared triangle, and $\varepsilon_v=\pm1$ is fixed by the chosen global orientation branch. Both $\theta_{vf}$ and $\xi_{vf}$ are smooth functions of the reconstructed four-simplex geometry, after a local choice of frame and spinor-phase gauge. The reconstructed four-simplex is insensitive to the $\SU(2)$ frame rotation at $e$ and to the $\mathrm U(1)$ phase of each spinor, since these redundancies act as rigid rotations of the tetrahedra or compensated frame changes.

\medskip

Let $f,g,h,k$ be the four triangular faces of a tetrahedron. We denote by
\begin{equation}
    \alpha^f_{gh}
\end{equation}
the intrinsic angle in $f$ between the two edges $f\cap g$, and $f\cap h$. The edge of $f$ opposite this angle is $f\cap k$.\footnote{Since $f$ is shared by two tetrahedra, the same triangle can equivalently be read off from the tetrahedron on the other side of $f$; there the two edges meet in the reversed cyclic order, giving $-\alpha^f_{gh}$, as expected for an oriented angle.} In terms of the spinors, the (oriented) 2D angle is given by
\begin{equation}
    \alpha^f_{gh}
    =
    \arg\left(
        \frac{
            \sbrasket{z_{vf}}{z_{vg}}\,
            \sbraket{z_{vf}}{z_{vg}}
        }{
            \sbrasket{z_{vf}}{z_{vh}}\,
            \sbraket{z_{vf}}{z_{vh}}
        }
    \right)\, .
    \label{eq:intrinsic-angle-spinors}
\end{equation}
Each bracket-product's phase tracks an edge direction relative to $z_{vf}$, so the ratio cancels this common reference and only leaves the angle between them \cite{Freidel:2013fia,Dona:2020yao}.
Geometric closure reconstructs a nondegenerate Euclidean tetrahedron. Hence each face is a nondegenerate Euclidean triangle, and its three exterior angles sum to $2\pi$. See the proof in Appendix~\ref{app:geometric-reconstruction-lemmas}.

\medskip

The squared length of the opposite edge $f\cap k$, reconstructed from the triangle $f$, is
\begin{equation}
    s_{f\cap k}^{(f)}
    \equiv
    \frac{
        2j_f\sin\alpha^f_{gh}
    }{
        \sin\alpha^f_{gk}\,
        \sin\alpha^f_{hk}
    }\, .
    \label{eq:reconstructed-edge-scalar}
\end{equation}
On the geometric constraint surface, this value is independent of the triangle, tetrahedron, and four-simplex used in its reconstruction. See the proof in Appendix~\ref{app:geometric-reconstruction-lemmas}.
Geometric closure reconstructs each tetrahedron from its areas and normals. Local flatness gives shape matching across every shared triangle inside a four-simplex. Shared dual-edge data identify the common tetrahedron of adjacent four-simplices. Assuming the star of each triangulation edge is connected, these equalities propagate through the star to every reconstruction of that edge. We may therefore define a single global scalar
\begin{equation}
    s_{f\cap k} \equiv s_{f\cap k}^{(f)}=s_{f\cap k}^{(k)} 
\end{equation}
for every triangulation edge.

\paragraph{The reconstruction map.}
To prove full equivalence with Regge calculus, rather than merely show that the Regge equations are among the constrained equations of motion, we must show that the reconstructed edge scalars form a complete set of local coordinates on the geometric constraint surface modulo gauge. The preceding construction defines a map
\begin{equation}
    R:
    \mathcal C_{\mathrm{geo}}
    \longrightarrow
    \mathcal{U}_{\mathrm{Regge}}\, ;
    \qquad
    R(q)=\{s_p\}\, .
    \label{eq:regge-reconstruction-map}
\end{equation}
Here, $\mathcal{U}_{\mathrm{Regge}}$ is the set of nondegenerate Lorentzian Regge geometries for which every four-simplex has spacelike boundary tetrahedra, on the fixed parity and globally consistent orientation branch chosen above. The fact that this map is well-defined is crucial, as it means that for each configuration in $\mathcal{C}_{\mathrm{geo}}$ there is exactly one nondegenerate Lorentzian Regge geometry. Furthermore, $R$ is gauge invariant, since neither an $\SU(2)$ frame rotation nor a spinor rephasing changes the reconstructed $s_p$ via \eqref{eq:reconstructed-edge-scalar}. Hence, $R$ descends to the quotient, defining a function on the orbit space
\begin{equation}
    \bar{R}:
    \mathcal C_{\mathrm{geo}}/\mathcal{G}
    \longrightarrow
    \mathcal{U}_{\mathrm{Regge}}\, ;
    \qquad
    \bar{R}([q])=R(q)=\{s_p\}\, .
\end{equation}
In addition, $\bar{R}$ is injective on the orbit space. If two constrained configurations reconstruct the same $\{s_p\}$, their tetrahedra have the same edge lengths. By rigidity of Euclidean tetrahedra (congruent tetrahedra of fixed parity are related by a unique rotation), their normals differ only by an $\SU(2)$ frame rotation. After aligning the normals, the corresponding spinors differ only by the $\mathrm U(1)$ rephasing redundancy. Since, on the chosen geometric branch, local flatness fixes every $\omega_{vf}$ in terms of the spinors, the two configurations differ only by a transformation in $\mathcal G$.

Conversely, let $r=\{s_p\}$ be an arbitrary point of $\mathcal{U}_{\mathrm{Regge}}$. Its edge lengths determine the area and shape of every triangle and tetrahedron. It therefore fixes the spins $j_f$ as the area of the triangles, and the outward unit normals $N_{ef}$. So it fixes a representative spinors $z_{ef}$ up to frames and spinor phases. It also fixes the oriented Lorentzian four-simplex at every vertex. The wedge angles are also fixed in terms of the spinors $\omega_{vf}=\varepsilon_v\theta_{vf}-i\xi_{vf}$ on the chosen orientation branch $\varepsilon_v$. As for the forward reconstruction, each of these steps is smooth on the nondegenerate locus: the areas, shape, and outward normals of a Euclidean tetrahedron depend smoothly on its six edge lengths, and the spinors and wedge angles depend smoothly on the normals and shape after the same local choice of frame and spinor-phase gauge used above. Therefore, given a point of $r$ we identified a point of $\mathcal{Q}$ up to gauge transformations. 

\begin{figure}[H]
\centering
\includegraphics[scale=0.8]{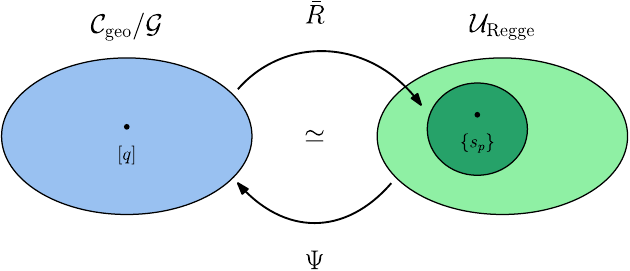}
\caption{\label{fig:reconstruction-diagram} The reconstruction map $\bar R$ sends the gauge class $[q]$ of a constrained configuration to the Regge geometry $\{s_p\}$ it reconstructs, and $\Psi$ builds a constrained configuration back from $\{s_p\}$. Since the two compositions are the identity on their respective spaces, $\bar R$ is a diffeomorphism on the chosen geometric branch.}
\end{figure}

The reconstructed normals of the tetrahedra satisfy geometric closure by construction, and the wedge holonomies, describing parallel transports inside a flat four-simplex, satisfy local flatness around every cycle. No further continuous physical holonomy data remain at fixed Regge edge lengths. Moreover, the phases $\omega_{ef}$ satisfies Eq.~\eqref{eq:phase-solution} since the Regge geometries in $\mathcal{U}_{\mathrm{Regge}}$ are Lorentzian by hypothesis. Hence, to any point $r=\{s_p\}$ in the space of nondegenerate Lorentzian Regge geometries with spacelike boundaries, we can associate (non-uniquely) a point in $\mathcal{C}_{\mathrm{geo}}$. After fixing the intrinsic tetrahedra and oriented four-simplices, the residual frame choices are precisely the elements of $\mathcal G$. This defines a map, shown together with $\bar{R}$ in Figure~\ref{fig:reconstruction-diagram},
\begin{equation}
    \Psi:
    \mathcal{U}_{\mathrm{Regge}}
    \longrightarrow
    \mathcal C_{\mathrm{geo}}/\mathcal G\, .
    \label{eq:inverse-regge-reconstruction}
\end{equation}
By construction, the image of a point under $\bar R$ is a Regge geometry with the prescribed areas and shapes. Therefore, $\Psi$ is a left inverse of $\bar R$:
\begin{equation}
    \Psi\circ\bar R
    =
    \operatorname{id}_{\mathcal C_{\mathrm{geo}}/\mathcal G}\, .
    \label{eq:left-inverse-reconstruction}
\end{equation}
In particular, $\Psi$ is surjective. We can also prove that $\Psi$ is injective. Suppose that $\Psi(r)=\Psi(r')$ for two Regge geometries $r,r'\in \mathcal{U}_{\mathrm{Regge}}$. Since their images belong to the same equivalence class in $\mathcal{C}_{\mathrm{geo}}$, the corresponding tetrahedra have the same face areas and outward normals, up to an overall rotation of each tetrahedron. The convex tetrahedra with given closed areas and normals is unique because of the Minkowski theorem. The corresponding tetrahedra are congruent and therefore have the same edge lengths, which implies $r=r'$. Hence, $\Psi$ is bijective and, since it is a left inverse of $\bar R$, it is also its right inverse:
\begin{equation}
    \bar R\circ\Psi
    =
    \operatorname{id}_{\mathcal{U}_{\mathrm{Regge}}}\, .
    \label{eq:right-inverse-reconstruction}
\end{equation}
Since both $\bar R$ and $\Psi$ are smooth by construction, they define mutually inverse diffeomorphisms on the chosen geometric branch.

The diffeomorphism statement has three consequences that will be essential below. First, every infinitesimal Regge variation of the edge scalars has a constraint-preserving lift to the spinfoam variables. Second, the edge variations $D_p$ form a complete basis of the physical tangent space $T_{[q]}(\mathcal C_{\mathrm{geo}}/\mathcal G)$, so stationarity along all $D_p$ is equivalent to stationarity along every constraint-preserving variation modulo gauge. Third, two constrained configurations with the same reconstructed edge scalars differ only by gauge, so no additional physical degrees of freedom remain at fixed Regge geometry. Thus the constrained variational problem contains neither missing Regge directions nor extra non-Regge directions.

\subsection{The action on the geometric constraint surface and its tangential variation}
\label{sec:action-geometric-surface}

We now evaluate the wedge action and its tangential variation on $\mathcal C_{\mathrm{geo}}$. On the constraint surface, the multiplier terms vanish. Consequently 
\begin{equation}
    S^{\EPRLclosed}\big|_{\mathcal C_{\mathrm{geo}}}=S^{\EPRLwedge}\big|_{\mathcal C_{\mathrm{geo}}}\, .
\end{equation}
Recall that, on the chosen globally consistent orientation branch, local flatness fixes $\omega_{vf}=\varepsilon_v\theta_{vf}-i\xi_{vf}$. With this convention the oriented deficit angle is
\begin{equation}
    \epsilon_f \equiv \sum_{v\subset f}\varepsilon_v\theta_{vf}\, ,
\end{equation}
and the twist terms cancel around an internal face. The action on $\mathcal C_{\mathrm{geo}}$ is $i$ times the Regge action in $\gamma$-spin-area units,
\begin{equation}
    S^{\EPRLclosed}\big|_{\mathcal C_{\mathrm{geo}}}
    =
    i\sum_fj_f\sum_{v\subset f}\left(\gamma\,\Re\omega_{vf}+\Im\omega_{vf}\right)
    =
    i\left[
    \sum_f\gamma j_f\epsilon_f - \sum_fj_f\sum_{v\subset f}\xi_{vf}
    \right]
    =
    iS_{\mathrm{Regge}}\, .
    \label{eq:wedge-action-on-geometric-sector}
\end{equation}

\medskip

Let us compute its variation on the geometric constraint surface. Let us identify the variation in a direction at a point $q$ with the tangent vector along that direction $D\in T_q Q$. The differential of any function $f$ in the direction of the variation $D$ can then be written as:
\begin{equation}
    df(D) = D f \, .
\end{equation}
Any constraint-preserving variation at a point $q\in\mathcal C_{\mathrm{geo}}$ is then a vector $D\in T_q\mathcal C_{\mathrm{geo}}$. By definition, it satisfies:
\begin{equation}
    dC^{\mathrm{geo}}_e(D)=0\,,
    \qquad
    dG_{vabc}(D)=0\,.
    \label{eq:linearized-geometric-constraints}
\end{equation}
The linearized geometric closure constraint $DC^{\mathrm{geo}}_e=0$ expresses the infinitesimal compatibility between the area variations and the variations of the spinorial normals, while $DG_{vabc}=0$ expresses the corresponding compatibility of the spinor and wedge-angle variations with local flatness and shape matching. These equations do not fix a variation uniquely: they leave untouched the spinor-phase redundancy and the common frame transformations at the tetrahedra, which we fix respectively by a phase condition and by any smooth local frame section.

Even after removing the gauge directions, there is no preferred basis of $T_{[q]}(\mathcal C_{\mathrm{geo}}/\mathcal{G})$. Any choice of local coordinates of $\mathcal C_{\mathrm{geo}}/\mathcal{G}$ gives a valid basis of the tangent space. A generic coordinate direction variation involves several reconstructed edge lengths at the same time and therefore produces a linear combination of the Regge equations. Since our goal is to reproduce the standard equation associated with each edge, we use the squared edge lengths $s_p$ as local coordinates and define the vector field $D_p$ as
\begin{equation}
    D_ps_{p'}=\delta_{pp'}\, .
\end{equation}
Thus $D_p$ changes $s_p$ while keeping all other squared edge lengths fixed. The differential of any function $\bar{f}$ on $\mathcal C_{\mathrm{geo}}/\mathcal{G}$ in the direction of the edge-coordinate variation $D_p$ can then be written as:
\begin{equation}
    \bar{\d}\bar{f}(D_p) = D_p \bar{f} \, .
\end{equation}
From now on, we restrict attention to the Regge variations $D_p$, with the goal of recovering the Regge equation associated with each edge $p$.

Notice that every tangent vector $D\in T_{[q]}(\mathcal C_{\mathrm{geo}}/\mathcal G)$ can be lifted nonuniquely to a vector
\begin{equation*}
    \tilde D\in T_q\mathcal C_{\mathrm{geo}} \, ,
\end{equation*}
such that $\d_q\pi(\tilde D)=D,$ with the ambiguity given by $\ker \d_q\pi$. This kernel consists precisely of the gauge directions; in particular,
\begin{equation*}
    \ker \d_q\pi\cong\operatorname{Lie}(\mathcal G)\, .
\end{equation*}
However, when varying a $\mathcal G$-invariant function, such as the action, any choice of lift gives the same variation as the corresponding variation on the orbit space. So, in particular, $D_p$ can be lifted, and
\begin{equation}
\tilde D_p S^{\EPRLclosed}\big|_{\mathcal C_{\mathrm{geo}}}
=
D_p\bar S^{\EPRLclosed}\big|_{\mathcal C_{\mathrm{geo}}}\, ,
\qquad
\forall\tilde D_p\in T_q\mathcal C_{\mathrm{geo}}
\text{ such that }\d\pi(\tilde D_p)=D_p\, .
\end{equation}
Therefore, from now on, we will not distinguish between the edge-coordinate variation and any of its lifts.

The Regge variation of the action requires only two projections of the complete tangent variation:
\begin{equation}
    D_p S^{\EPRLclosed}\big|_{\mathcal C_{\mathrm{geo}}} = iD_p \sum_fj_f\sum_{v\subset f}\left(\gamma\,\Re\omega_{vf} +\Im\omega_{vf} \right)\, ,
    \label{eq:action-tangential-variation}
\end{equation}
where we used \eqref{eq:linearized-geometric-constraints}. Using the linearity of $D_p$ and Leibniz's rule, to compute the Regge variation of the action, we need only the two fundamental variations
\begin{equation}
    D_p j_f\,,
    \qquad
    \text{and}
    \qquad
    \sum_{f\subset v} j_f D_p \omega_{vf}\,.
    \label{eq:required-regge-projections}
\end{equation}
The next two subsections determine these quantities without constructing the remaining components of $D_p$ explicitly since their expression is not needed and unreadable (even if in line of principle is computable from \eqref{eq:linearized-geometric-constraints}).

\subsection{Spin component of the Regge variations}
\label{sec:spin-component-regge-variations}
\begin{figure}[H]
\centering
\includegraphics[scale=0.9]{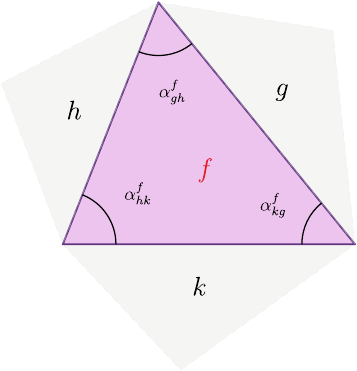}
\caption{\label{fig:triangle-angles} The triangle $f$ and its three neighbors $g,h,k$ within the same tetrahedron, drawn as the unfolded net so that each neighbor sits across the edge it shares with $f$. The intrinsic angle $\alpha^f_{gh}$ is the one opposite the edge $f\cap k$, and similarly for $\alpha^f_{hk}$ and $\alpha^f_{kg}$.}
\end{figure}

Consider a triangle $f$ belonging to a tetrahedron with faces $f,g,h,k$ (Figure~\ref{fig:triangle-angles}), with $2$D angles $\alpha^f_{gh}$, $\alpha^f_{hk}$, $\alpha^f_{kg}$ opposite to the edges $f\cap k$, $f\cap g$, $f\cap h$ respectively, satisfying
\begin{equation}
    \alpha^f_{gh}+\alpha^f_{hk}+\alpha^f_{kg}=2\pi\, .
    \label{eq:triangle-angle-sum}
\end{equation}
The three reconstructed squared edge lengths are
\begin{align}
s_{f\cap k} =
\frac{2 j_f\sin\alpha^f_{gh}}{\sin\alpha^f_{hk}\sin\alpha^f_{kg}}\,, \quad
s_{f\cap g} =
\frac{2 j_f\sin\alpha^f_{hk}}{\sin\alpha^f_{kg}\sin\alpha^f_{gh}}\,, \quad
s_{f\cap h} =
\frac{2 j_f\sin\alpha^f_{kg}}{\sin\alpha^f_{gh}\sin\alpha^f_{hk}}\,.
\label{eq:three-reconstructed-lengths}
\end{align}
Let $D_{f\cap k}$ be the Regge variation associated with the edge $f\cap k$, so that by definition
\begin{equation}
    D_{f\cap k}s_{f\cap k}=1\, ,
    \qquad
    D_{f\cap k}s_{f\cap g} =0\, , 
    \qquad \text{and}\qquad
    D_{f\cap k}s_{f\cap h}=0\, .
    \label{eq:local-regge-coordinate-variation}
\end{equation}

\medskip

Taking logarithmic derivatives of the ratios $s_{f\cap k}/s_{f\cap g}$ and $s_{f\cap k}/s_{f\cap h}$ in \eqref{eq:three-reconstructed-lengths}, and using \eqref{eq:local-regge-coordinate-variation}, gives
\begin{equation}
    \cot\alpha^f_{gh}\,D_{f\cap k}\alpha^f_{gh}
    -
    \cot\alpha^f_{hk}\,D_{f\cap k}\alpha^f_{hk}
    =
    \cot\alpha^f_{gh}\,D_{f\cap k}\alpha^f_{gh}
    -
    \cot\alpha^f_{kg}\,D_{f\cap k}\alpha^f_{kg}
    =
    \frac{1}{2s_{f\cap k}}\, .
    \label{eq:length-variation-ab}
\end{equation}
The two equalities imply
\begin{equation}
    \cot\alpha^f_{hk}\,D_{f\cap k}\alpha^f_{hk}
    =
    \cot\alpha^f_{kg}\,D_{f\cap k}\alpha^f_{kg}\, ,
    \label{eq:fixed-side-angle-relation}
\end{equation}
while the product $s_{f\cap g}s_{f\cap h}=4j_f^2/\sin^2\alpha^f_{gh}$ is fixed under $D_{f\cap k}$, giving
\begin{equation}
    \frac{D_{f\cap k}j_f}{j_f}
    =
    \cot\alpha^f_{gh}\,D_{f\cap k}\alpha^f_{gh}\, .
    \label{eq:area-angle-variation}
\end{equation}
On the other hand, differentiating \eqref{eq:triangle-angle-sum} and eliminating $D_{f\cap k}\alpha^f_{kg}$ with \eqref{eq:fixed-side-angle-relation} gives, after multiplying by $\cos\alpha^f_{gh}\cos\alpha^f_{hk}\cos\alpha^f_{kg}$ and simplifying with the angle-sum identity,
\begin{equation}
    \cot\alpha^f_{hk}\,D_{f\cap k}\alpha^f_{hk}\,\cos\alpha^f_{gh}
    =
    \cot\alpha^f_{gh}\,D_{f\cap k}\alpha^f_{gh}\,\cos\alpha^f_{hk}\cos\alpha^f_{kg}\, .
    \label{eq:angle-variation-ab}
\end{equation}
Multiplying \eqref{eq:length-variation-ab} by $\cos\alpha^f_{gh}$, using \eqref{eq:angle-variation-ab}, and simplifying with the angle-sum identity $\cos\alpha^f_{gh}=\cos\alpha^f_{hk}\cos\alpha^f_{kg}-\sin\alpha^f_{hk}\sin\alpha^f_{kg}$ gives
\begin{equation}
    \cot\alpha^f_{gh}\,D_{f\cap k}\alpha^f_{gh}
    =
    -\frac{1}{2s_{f\cap k}}\,
    \frac{\cos\alpha^f_{gh}}
    {\sin\alpha^f_{hk}\sin\alpha^f_{kg}}\, .
    \label{eq:angle-solved}
\end{equation}
Multiplying \eqref{eq:angle-solved} by $j_f\tan\alpha^f_{gh}$ and using \eqref{eq:three-reconstructed-lengths} to write $s_{f\cap k}\sin\alpha^f_{hk}\sin\alpha^f_{kg}=2j_f\sin\alpha^f_{gh}$ gives the clean intermediate result
\begin{equation}
    j_f\,D_{f\cap k}\alpha^f_{gh}
    =
    -\frac14\, .
    \label{eq:angle-variation-clean}
\end{equation}
Combined with \eqref{eq:area-angle-variation}, this gives
\begin{equation}
    D_{f\cap k}j_f
    =
    \cot\alpha^f_{gh}\,\left(j_f\,D_{f\cap k}\alpha^f_{gh}\right)
    =
    -\frac14\cot\alpha^f_{gh}\, .
\end{equation}
Restoring the general edge/face notation, and remembering that the variation is null unless $p$ belongs to the face $f$
\begin{equation}
    D_pj_f
    =
    \begin{cases}
        -\frac14\,
        \cot\alpha_{fp} & p\subset f\\
        0 & p\not\subset f
    \end{cases}\, .
        \label{eq:spin-component-regge-variation}
\end{equation}

\subsection{Weighted wedge-angle variation - Scl\"afli indentity}
\label{sec:weighted-wedge-angle-variation}
We now compute the combination
\begin{equation}
    \sum_{f\subset v}j_fD_p\omega_{vf} \, ,
\end{equation}
without determining the individual variations $D_p\omega_{vf}$. We consider the fixed spinor phases (this is a gauge choice), the variations of the normalized spinors $\ket{z_{vf}}$ are then proportional to their dual
\begin{equation}
    \ket{D_pz_{vf}}
    =
    \beta_{vf}^{(p)}\sket{z_{vf}}\,.
    \label{eq:phase-fixed-spinor-variation}
\end{equation}
The complex coefficient $\beta_{vf}^{(p)}$ could be determined from the linearized geometric closure constraint using the knowledge of $D_pj_f$, but we do not need its explicit value. Fix an edge $vf$, the spinorial parametrization gives
\begin{equation}
    g_{vf}\ket{z_{vf}}
    =
    e^{\omega_{vf}/2}\sket{\tilde z_{vf}}\, ,
\end{equation}
Taking the variation of this equation, using \eqref{eq:phase-fixed-spinor-variation} and the Leibniz rule gives
\begin{equation}
    (D_p g_{vf})\ket{z_{vf}} + g_{vf}\beta_{vf}^{(p)}\sket{z_{vf}}
    =
    \frac{1}{2}(D_p\omega_{vf})e^{\omega_{vf}/2}\sket{\tilde z_{vf}} - e^{\omega_{vf}/2}\overline{\tilde\beta_{vf}^{(p)}}\ket{\tilde z_{vf}}\, .
\end{equation}
Acting with $g_{vf}^{-1}$ on the left and using the spinor inner products, we find that
\begin{equation}
    g_{vf}^{-1}(D_p g_{vf})\ket{z_{vf}} + \beta_{vf}^{(p)}\sket{z_{vf}}
    =
    \frac{1}{2}(D_p\omega_{vf})\ket{z_{vf}} + e^{\omega_{vf}}\overline{\tilde\beta_{vf}^{(p)}}\sket{z_{vf}}\, .
\end{equation}
Projecting on $\bra{z_{vf}}$ gives 
\begin{equation}
    \bra{z_{vf}}
    g_{vf}^{-1}D_pg_{vf}
    \ket{z_{vf}}
    =
    \frac{1}{2}D_p\omega_{vf}\,,
\end{equation}
and analogously 
\begin{equation}
    \sbra{z_{vf}}
    g_{vf}^{-1}D_pg_{vf}
    \sket{z_{vf}}
    =
    -\frac12D_p\omega_{vf}\,.
\end{equation}
Using the normal generator of Appendix~\ref{app:spinors},
\begin{equation}
    N_{vf}
    =
    \sket{z_{vf}}\sbra{z_{vf}}
    -
    \ket{z_{vf}}\bra{z_{vf}}\, ,
\end{equation}
we obtain
\begin{equation}
    \Tr\left(
        N_{vf}g_{vf}^{-1}D_pg_{vf}
    \right)
    =
    -D_p\omega_{vf}\, .
    \label{eq:wedge-angle-trace-identity}
\end{equation}

We now use linearized local flatness. Choose one tetrahedron $r$ as reference. Local flatness on the cycle $r\to a\to b\to r$ gives
\begin{equation}
    g_{v f_{rb}}\,
    g_{v f_{ba}}\,
    g_{v f_{ar}}
    =
    \Id\, .
\end{equation}
Since $g_{v f_{rb}}=g_{v f_{br}}^{-1}$, it follows that
\begin{equation}
    g_{v f_{ba}}
    =
    g_{v f_{br}}\,
    g_{v f_{ar}}^{-1}\, .
\end{equation}
Taking the variation of this equation gives,
\begin{equation}
    D_pg_{v f_{ba}}
    = D_pg_{v f_{br}}\,g_{v f_{ar}}^{-1}
    -
    g_{v f_{br}}g_{v f_{ar}}^{-1}
    D_pg_{v f_{ar}}\,g_{v f_{ar}}^{-1}\, .
\end{equation}
Using $g_{v f_{ba}}^{-1} = g_{v f_{ar}}g_{v f_{br}}^{-1}$, we obtain
\begin{equation}
    g_{v f_{ba}}^{-1}D_pg_{v f_{ba}} = g_{v f_{ar}}g_{v f_{br}}^{-1}
    D_pg_{v f_{br}}\,g_{v f_{ar}}^{-1} - D_pg_{v f_{ar}}\,g_{v f_{ar}}^{-1}\, .
\end{equation}
The wedge-angle variation is therefore
\begin{equation}
    D_p\omega_{v f_{ba}}
    =
    -\Tr\left(
        N_{v f_{ba}}\,
        g_{v f_{ar}}g_{v f_{br}}^{-1}
        D_pg_{v f_{br}}\,g_{v f_{ar}}^{-1}
    \right)
+
    \Tr\left(
        N_{v f_{ba}}\,
        D_pg_{v f_{ar}}\,g_{v f_{ar}}^{-1}
    \right)\, .
\end{equation}
The source and target normals of the wedge satisfy
\begin{equation}
    g_{v f_{ba}}\,
    N_{v f_{ba}}\,
    g_{v f_{ba}}^{-1}
    =
    -N_{v f_{ab}}\, .
\end{equation}
Using $g_{v f_{ba}}=g_{v f_{br}}g_{v f_{ar}}^{-1}$, this becomes
\begin{equation}
    g_{v f_{ar}}^{-1}
    N_{v f_{ba}}
    g_{v f_{ar}}
    =
    -
    g_{v f_{br}}^{-1}
    N_{v f_{ab}}
    g_{v f_{br}}\, .
\end{equation}
Cyclicity of the trace then gives
\begin{equation}
        D_p\omega_{v f_{ba}}
        =
        \Tr\left(
            N_{v f_{ab}}\,
            D_pg_{v f_{br}}\,g_{v f_{br}}^{-1}
        \right)
        +
        \Tr\left(
            N_{v f_{ba}}\,
            D_pg_{v f_{ar}}\,g_{v f_{ar}}^{-1}
        \right)\, .
\end{equation}
Moving to the edge notation, let $f$ denote the face underlying the wedge $vf_{ba}$. Since the wedge is oriented from $a$ to $b$, its source and target normals are
\begin{equation}
    N_{v f_{ba}}=N_{af}\, ,
    \qquad
    N_{v f_{ab}}=N_{bf}\, .
\end{equation}
The two terms are therefore associated with the closure equations at edges $b$ and $a$, respectively. Multiplying by $j_f$, summing over all faces at $v$, and regrouping the terms by edge gives
\begin{align}
    \sum_{f\subset v}j_fD_p\omega_{vf}
    &=
    \sum_{a\neq r}
    \Tr\left(
        \Big(
            \sum_{\substack{f\subset v\\f\ni a}}
            j_fN_{af}
        \Big)
        D_pg_{v f_{ar}}\,g_{v f_{ar}}^{-1}
    \right)
    \nonumber
    \\
    &=
    \sum_{a\neq r}
    \Tr\left(
        C_a^{\mathrm{geo}}\,
        D_pg_{v f_{ar}}\,g_{v f_{ar}}^{-1}
    \right)=0\, .
\end{align}
In the second equality we used the fact that the four faces at $v$ containing the edge $a$ are precisely the four faces entering the geometric-closure constraint
\begin{equation}
    C_a^{\mathrm{geo}}
    =
    \sum_{f\ni a}j_fN_{af}
    =
    0\, .
\end{equation}
Hence
\begin{equation}
    \sum_{f\subset v}j_fD_p\omega_{vf}=0\, .
    \label{eq:weighted-wedge-angle-vanishes}
\end{equation}

\subsection{The Regge equations of motion}
\label{sec:regge-eom}
By constrained stationarity, $D_pS^{\EPRLclosed}\big|_{\mathcal C_{\mathrm{geo}}}=0$ for every internal triangulation edge $p$. Applying Leibniz's rule to \eqref{eq:action-tangential-variation} gives
\begin{equation}
    i\left[
    \sum_fD_pj_f\sum_{v\subset f}\left(\gamma\,\Re\omega_{vf}+\Im\omega_{vf}\right)
    +
    \sum_fj_f\sum_{v\subset f}\left(\gamma\,\Re D_p\omega_{vf}+\Im D_p\omega_{vf}\right)
    \right] = 0\, .
\end{equation}
Reordering the second double sum by vertex and using \eqref{eq:weighted-wedge-angle-vanishes},
\begin{equation}
    \sum_fj_f\sum_{v\subset f}\left(\gamma\,\Re D_p\omega_{vf}+\Im D_p\omega_{vf}\right)
    =
    \sum_v\left[
        \gamma\,\Re\!\left(\sum_{f\subset v}j_fD_p\omega_{vf}\right)
        +
        \Im\!\left(\sum_{f\subset v}j_fD_p\omega_{vf}\right)
    \right]
    =0\, ,
\end{equation}
so only the first term survives. Since the twist terms cancel around every internal face, $\sum_{v\subset f}\Im\omega_{vf}=0$ \eqref{eq:wedge-action-on-geometric-sector}, and the surviving coefficient reduces to $\gamma\epsilon_f$, giving
\begin{equation}
    \gamma\sum_fD_pj_f\,\epsilon_f = 0\, .
\end{equation}
Using \eqref{eq:spin-component-regge-variation}, $D_pj_f=-\frac14\cot\alpha_{fp}$ if $p\subset f$, this becomes
\begin{equation}
    \sum_{f\supset p}\epsilon_f\cot\alpha_{fp}=0 \, ,
    \qquad
    \text{for every internal triangulation edge }p\, .
    \label{eq:regge-eom-final}
\end{equation}
This is the classical Regge equation of motion for the edge variable $s_p$ \cite{Regge:1961px}: at every internal edge, the deficit-angle-weighted variation of the areas of the incident triangles vanishes.

On a regular, nondegenerate, consistently oriented Lorentzian geometric component with spacelike tetrahedra, the gauge classes of stationary points of the EPRL-FK wedge action subject to local flatness and geometric closure are therefore locally in one-to-one correspondence with solutions of the length-Regge equations for the internal edge variables. The quotient by $\mathcal G$ ensures that gauge-equivalent wedge configurations do not represent distinct physical solutions. In this constrained variational problem, the independent spin variations are replaced by Regge variations tangent to the geometric constraint surface, and the flatness equation is replaced by the Regge equations.

\section{A candidate implementation of geometric closure in the EPRL-FK amplitude}
\label{sec:changing-eprl-fk-edge}
The geometric closure constraint cannot be implemented using the standard EPRL-FK intertwiner alone. In the standard amplitude, the wedges meeting at a tetrahedron are glued together by an intertwiner. This is usually written as a coherent intertwiner \cite{Livine:2007vk}: a tensor product of one $\SU(2)$ coherent state for each face of the tetrahedron, projected onto the invariant subspace by group averaging. Because of the $\SU(2)$ invariance of the vertex amplitude, this group average can be reabsorbed into a change of variables in the wedge-holonomy integrals. What remains are the spinors of the coherent states gluing consecutive wedges together. This is the usual construction leading to the EPRL-FK amplitude, reviewed in Appendix~\ref{app:vertex-amplitude-details}.

The result is precisely the action \eqref{eq:EPRL-wedge}. The coherent intertwiner satisfies the quantum Gauss constraint, but this does not impose geometric closure on the wedge holonomies off shell. In particular, it does not tie the spins $j_f$ to the normals $N_{ef}$ extracted from those holonomies.

A first attempt would be to use only coherent intertwiners whose labels satisfy closure, i.e. to restrict the integral over the spinors to those satisfying the geometric closure constraint. This does not work. It has been shown that coherent intertwiners labelled by exactly closed configurations still give a resolution of the identity on the intertwiner space, provided they are integrated with the appropriate measure \cite{Conrady:2009px}. Replacing the usual coherent-intertwiner resolution with this closed resolution therefore leaves the amplitude unchanged. It changes the parametrization of the intertwiner space, but it does not add a new constraint to the model. In particular, the wedge holonomies are still not required to satisfy geometric closure.

A genuine modification must act directly on the relation between the spins and the wedge-holonomy data. The most immediate possibility is to insert
\begin{equation}
\prod_e
\delta^{(3)}
\left(
\sum_{f\ni e}j_fN_{ef}
\right)
\end{equation}
in the wedge-holonomy representation of the amplitude. Exponentiating these delta functions with one multiplier $\lambda_e$ for each edge produces
\begin{equation}
i\sum_e
\Tr\left(
\lambda_e
\sum_{f\ni e}j_fN_{ef}
\right)\, ,
\end{equation}
which is exactly the additional term used in the constrained action studied above. This gives a first candidate implementation of geometric closure at the amplitude level. It looks local at the edge, since it only couples the four spins and spinors belonging to the same tetrahedron. It is not, however, a modification of the usual edge amplitude $A_e(i_e)$: it is a new gluing factor depending simultaneously on the spins and on the wedge-holonomy data.

The normal $N_{ef}$ can be expressed in terms of the wedge holonomy using the polar decomposition of $\SL(2,\C)$ matrices. In particular, the product of a wedge holonomy with its adjoint contains only the information about the source face normal and the boost angle:
\begin{equation}
g_{vf}^\dagger g_{vf}
= e^{\Re \omega_{vf}} \ket{z_{vf}}\bra{z_{vf}}
+ e^{-\Re \omega_{vf}} \sket{z_{vf}}\sbra{z_{vf}}
= e^{-\Re \omega_{vf} N_{ef}} \, ,
\end{equation}
where $g_{vf}$ is the wedge holonomy associated with the edge $e$ and the vertex $v$. Therefore, away from the vanishing-boost locus and after choosing the sign branch $\sigma_{vf}\equiv\operatorname{sgn}(\Re\omega_{vf})$, we can isolate the normal $N_{ef}$ from the wedge holonomy as
\begin{equation}
L_{vf}\equiv\log\left(g_{vf}^\dagger g_{vf}\right)\, ,
\qquad
N_{ef}
=-\frac{L_{vf}}{\Re\omega_{vf}}
=-\sigma_{vf}\frac{L_{vf}}{\sqrt{\frac12\Tr(L_{vf}^{2})}}\, .
\end{equation}
Inserting it in the closure delta function gives a candidate implementation of geometric closure in terms of the wedge holonomies. 
\begin{equation}
-i\sum_e
\Tr\left(
\lambda_e
\sum_{f\ni e}j_f
\sigma_{vf}
\frac{L_{vf}}{\sqrt{\frac12\Tr(L_{vf}^{2})}}
\right)\, ,
\end{equation}
where $g_{vf}$ is the wedge holonomy whose source is the edge $e$, and $\sigma_{vf}$ is part of the chosen logarithm branch. Reversing this branch reverses the extracted normal. This form also makes clear that the holonomy-only expression is ill defined at critical points with $\Re\omega_{vf}=0$. We do not know whether this is a defect or a useful feature: such configurations lie outside the nondegenerate geometric region used in the theorem of Section~\ref{sec:closure-spinors}. The important caveat is that, while ``local'' at the edge, this expression still depends on the wedge holonomies associated with all the wedges meeting at the edge, and therefore on data from both vertices surrounding the edge. This is a nonlocal modification of the amplitude, and it is not clear whether it can be implemented in a way that preserves the factorization of the amplitude into vertex contributions.

\section{Discussion}
\label{sec:discussion}
The purpose of this paper was to isolate, in spinfoam variables, the precise condition that turns the local Regge geometries selected by the EPRL-FK vertex into Regge dynamics on a fixed simplicial complex. In the wedge-holonomy parametrization, local flatness already has a clear geometric meaning: it is the shape-matching condition that lets the spinorial data around a vertex reconstruct a Lorentzian four-simplex. What it does not do is tie the face spins to the areas of the tetrahedra reconstructed from those spinors. The missing condition is geometric closure, imposed as a strong holonomy-flux compatibility constraint at each tetrahedron. On the regular nondegenerate Lorentzian branch with spacelike boundary tetrahedra considered here, local flatness together with geometric closure defines a constraint surface equivalent, modulo gauge, to the space of length-Regge geometries. The EPRL-FK wedge action restricted to this surface then gives the Regge equations instead of the flatness equation.

This gives a concrete diagnosis of the flatness problem in the wedge description. In the standard semiclassical variation, the spins are independent variables and the action is linear in them. Once the local-flatness equations have reconstructed the deficit angle around a face, varying the spin sets this angle to zero. This is the mechanism behind the usual flatness equation found in the asymptotic analysis of the EPRL-FK amplitude \cite{Conrady:2008mk,Bonzom:2009hw,Hellmann:2013gva}. In the geometrically closed problem the spin variation is no longer an arbitrary area variation: it must be tangent to the Regge constraint surface. The Schlaefli identity then removes the variations of the dihedral angles and the remaining equations are the length-Regge equations.

The result is close in spirit to area-angle Regge calculus \cite{Dittrich:2008va,Dittrich:2008ar}. There too, areas and three-dimensional angles form a larger configuration space than edge lengths, and Regge calculus is recovered only after imposing both shape matching and closure. The present work translates this statement into the wedge variables of the EPRL-FK amplitude. Local flatness plays the role of shape matching, while geometric closure is the additional condition that makes the spins compatible with the tetrahedral normals encoded by the holonomies. In this sense, the constrained wedge system is the spinfoam analogue of the area-angle Regge variational problem restricted to the length-Regge sector.

This should also be distinguished from the use of closed coherent intertwiners. The coherent intertwiner, or the quantum Gauss constraint on the intertwiner space, imposes gauge invariance at the kinematical level. Even restricting the coherent-intertwiner labels to closed configurations does not modify the amplitude, because closed coherent intertwiners still provide a resolution of the identity with the appropriate measure \cite{Conrady:2009px}. Geometric closure in the present sense is stronger and more off shell: it constrains the spins together with the normals extracted from the wedge holonomies entering the amplitude. It is therefore not a change of basis in the intertwiner space, but a new compatibility condition on the semiclassical integration variables.

Related concerns about the status of closure in spin-foam models have appeared previously \cite{Alexandrov:2007pq}. In that context, coherent-state constructions were observed not to impose classical closure exactly on their labels, even though closed configurations dominate semiclassically. Subsequent analyses emphasized instead that strong classical closure may be regarded as an on-shell equation of motion, while the off-shell quantum condition takes a relaxed, normal-dependent covariant form \cite{Alexandrov:2008da,Alexandrov:2010un,Alexandrov:2012pj}. The constraint studied here is different from both this relaxed closure and the ordinary Gauss constraint: it promotes the specific holonomy--flux compatibility relation from an on-shell property of geometric critical configurations to an independent constraint of the semiclassical variational problem. As explained in Section~\ref{sec:dynamical-vs-strong}, this changes the allowed spin variations into variations tangent to the Regge constraint surface.

There is also a useful comparison with the program based on complex critical points. Curved Regge geometries can be recovered as complex saddles of the standard EPRL-FK action, and their effective action reproduces the Regge equations in a suitable small-$\gamma$ regime \cite{Han:2021kll,Han:2023cen,Han:2024lti}. Our proposal is different. We do not look for curved Regge geometries away from the real critical surface of the original action. Instead, we modify the real constrained variational problem so that its geometric critical surface is already the Regge one. The two perspectives may be complementary: complex saddles describe how curvature appears in the original amplitude, while geometric closure asks what extra condition would make the real semiclassical variational principle reproduce Regge calculus directly.

The comparison with continuum-limit analyses is similarly suggestive. On regular lattices, the semiclassical spinfoam dynamics has been argued to reproduce the graviton dynamics of General Relativity at leading order, while its subleading behavior points to additional non-length degrees of freedom naturally described by an area metric \cite{Dittrich:2022yoo,Borissova:2022clg,Borissova:2023yxs}. Recent work has also given a microscopic version of this statement by identifying twisted four-simplex data with cyclic area-metric data \cite{Dittrich:2023rcr}. In that language the extra degrees of freedom are shape-mismatching, or more generally non-metric, directions rather than length-Regge ones. Our result is compatible with this picture but addresses a different constraint: once local flatness has imposed shape matching, the remaining non-Regge freedom in the wedge variables is the possible failure of the spins to close with the normals reconstructed from the holonomies. Imposing geometric closure removes these non-closed directions locally, before taking a continuum limit. Whether this strong restriction is the right microscopic modification, or whether the extra area-metric modes should instead be kept and controlled dynamically or through renormalization, remains an open question.

At the amplitude level, the most direct implementation is to insert a closure delta function at each edge. In spinor variables this is local at the tetrahedron, but it is not simply a modification of the usual numerical edge amplitude: it couples the spins to the wedge-holonomy data. Rewriting the same condition purely in terms of $g_{vf}$ makes the proposal more intrinsic, but also exposes two unresolved issues. First, it depends on all wedge holonomies meeting at an edge and therefore does not obviously preserve the usual factorization into vertex contributions. Second, the expression is ill defined when $\Re\omega_{vf}=0$. This may be a pathology, or it may be harmless, since such configurations lie outside the nondegenerate geometric region used in our construction. A complete quantum definition would have to decide how these sectors are treated.

Several limitations should be kept explicit. The equivalence with Regge calculus is local on a regular component of the constraint surface. It assumes nondegenerate Lorentzian four-simplices with spacelike boundary tetrahedra, consistent orientation, connected edge stars, and a fixed branch of the reconstruction. We have not constructed a full quantum measure for the closure insertion, proved uniqueness of the modification, or analyzed degenerate and orientation-changing sectors. We have also treated the spins as continuous large-spin variables, as appropriate for the stationary-phase analysis, and have not addressed the effects of spin discreteness or the behavior of the spin sum.

We have not analyzed how the simplicity constraints are imposed in the EPRL-FK model beyond the assumptions entering the wedge action. In particular, the derivation of $S^{\EPRLwedge}$ in Appendix~\ref{app:vertex-amplitude-details} requires the dominant-saddle condition. Giving up this condition would require a different implementation of the simplicity constraints to recover Regge dynamics and lies outside the scope of the current analysis.

The difficulties encountered by the holonomy-only expression suggest a more radical possibility. Rather than taking the $\SL(2,\C)$ wedge holonomies as independent integration variables, one could formulate the amplitude directly in terms of the spinors $z_{ef}$ and complex wedge angles $\omega_{vf}$, reconstructing $g_{vf}$ from them through \eqref{eq:wedge-param}. This would follow the philosophy of the holomorphic and twistorial formulations of spinfoams, in which amplitudes are expressed as integrals over spinorial variables and the sums over representation data are encoded through homogeneity or contour variables \cite{Dupuis:2011fz,Banburski:2014cwa,Hnybida:2015ioa,Speziale:2012nu}. In such a formulation, geometric closure would remain the manifestly local and regular condition $\sum_{f\ni e}j_fN(z_{ef})=0$, while local flatness would be imposed on the wedge holonomies reconstructed from the spinors and complex angles. This could avoid extracting the normals through a polar decomposition, and hence the singularity at vanishing boost, while making the holonomy--flux compatibility condition local at the tetrahedron. 

The main lesson is therefore conditional but sharp. If one asks for a real semiclassical spinfoam variational principle whose geometric sector is equivalent to length-Regge calculus, local flatness is not enough. One must also impose holonomy-flux compatibility between the spins and the normals reconstructed from the wedge holonomies. With that extra condition, the flatness equation is replaced by the Regge equations on the nondegenerate geometric branch.

\section*{Acknowledgements}
We thank Muxin Han, Simone Speziale, and Eugenio Bianchi for useful discussions during the early stages of this work at the WOST kick-off meeting in Verona. We also thank Carlo Rovelli and Wolfgang Wieland for their feedback on the first version of this manuscript. This work was made possible through the support of the WOST, \href{https://withoutspacetime.org}{WithOut SpaceTime project}, supported by Grant ID~\#63683 from the John Templeton Foundation. M.B. acknowledges support by Grant~\#63132 from the John Templeton Foundation, as part of the \href{https://eff.cstq.org}{Enrico Fermi Fellowship program}. The opinions expressed in this publication are those of the authors and do not necessarily reflect the views of the respective funding body.

\appendix

\section{Spinors, conventions and all that}
\label{app:spinors}

A spinor is given by a couple of complex numbers $\ket{z} = (z^0, z^1) \in \C^2$, with bra $\bra{z} \equiv \ket{z}^\dagger$. The scalar product of two spinors $\ket{z}$ and $\ket{w}$ is defined by
\begin{equation}
    \braket{z}{w} = \bar z^0 w^0 + \bar z^1 w^1 \, ,
    \label{eq:app-sp-brackets}
\end{equation}
Its norm is $\braket{z}{z} = |z^0|^2 + |z^1|^2$, and we call $\ket{z}$ a \emph{unit spinor} whenever $\braket{z}{z}=1$. The dual spinor is
\begin{equation}
    \sket{z} \equiv (-\bar z^1, \bar z^0) \, ,
    \label{eq:app-sp-dual}
\end{equation}
with bra $\sbra{z}\equiv \sket{z}^\dagger$, from which follows immediately that the dual spinor is orthogonal to the original, $\sbraket{z}{z}=0$. Therefore, a unit spinor $\ket{z}$ and its dual $\sket{z}$ form an orthonormal basis of $\C^2$. The four independent products of two spinors $\ket{z}$ and $\ket{w}$ are
\begin{equation}
    \braket{z}{w} \, , \qquad
    \sbrasket{z}{w} \, \qquad
    \sbraket{z}{w} \, , \qquad
    \brasket{z}{w} \, .
    \label{eq:app-sp-products}
\end{equation}
The first two products are complex conjugates of each other, and similarly for the last two
\begin{equation}
    \sbrasket{z}{w} = \braket{w}{z} = \overline{\braket{z}{w}} \, , \qquad
    \brasket{z}{w} = - \brasket{w}{z} = -\overline{\sbraket{z}{w}} \, .
    \label{eq:app-sp-conjugation}
\end{equation}
Finally, the sum of the projectors onto a unit spinor and its dual gives the identity on $\C^2$,
\begin{equation}
    \ket{z}\bra{z} + \sket{z}\sbra{z} = \Id \, .
    \label{eq:app-sp-identity}
\end{equation}
From which follows the Lagrange identity
\begin{equation}
    \braket{w}{z}\braket{z}{v} + \brasket{w}{z}\sbraket{z}{v} = \braket{w}{v} \, .
    \label{eq:app-sp-lagrange}
\end{equation}

Every unit spinor determines a point on the sphere,
\begin{equation}
    \vec n(z) \equiv -\bra{z}\vec\sigma\ket{z} \, , \qquad |\vec n(z)|=1 \, ,
    \label{eq:app-sp-normal}
\end{equation}
with $\vec\sigma=(\sigma_1,\sigma_2,\sigma_3)$ the Pauli matrices. The dual spinor identify the same direction on the sphere with opposite verse
\begin{equation*}
    \vec n(z) \equiv \sbra{z}\vec\sigma\sket{z} \, , \qquad |\vec n(z)|=1 \, .
\end{equation*}
All the outer products used in the main text follow from a single fact: since $\{\Id,\vec\sigma\}$ is a basis of $2\times2$ complex matrices, any two spinors $u,v$ satisfy the Fierz identity
\begin{equation}
    \ket{u}\bra{v} = \frac12\Big(\braket{v}{u}\,\Id + \bra{v}\vec\sigma\ket{u}\,\cdot\vec\sigma\Big) \, .
    \label{eq:app-sp-fierz}
\end{equation}
For a unit spinor and its dual implies
\begin{equation}
    \ket{z}\bra{z} = \frac{\Id - \vec n(z)\cdot\vec\sigma}{2}\, , \quad \text{and} \quad     \sket{z}\sbra{z} = \frac{\Id + \vec n(z)\cdot\vec\sigma}{2}\, ,
\end{equation}
Together these reproduce again the resolution of the identity and the traceless generator used throughout the paper,
\begin{equation}
    \ket{z}\bra{z} + \sket{z}\sbra{z} = \Id \, , \qquad
    N(z) \equiv \sket{z}\sbra{z} - \ket{z}\bra{z} = \vec n(z) \cdot \vec\sigma \, .
    \label{eq:app-sp-resolution}
\end{equation}

For two unit spinors, the Fierz identity \eqref{eq:app-sp-fierz} gives
\begin{equation}
    |\braket{z}{w}|^2 = \cos^2\!\Big(\frac{\theta}{2}\Big) = \frac{1+\vec n(z)\cdot\vec n(w)}{2} \, , \qquad
    |\sbraket{z}{w}|^2 = \sin^2\!\Big(\frac{\theta}{2}\Big) = \frac{1-\vec n(z)\cdot\vec n(w)}{2} \, ,
    \label{eq:app-sp-angle}
\end{equation}
with $\theta$ the angle between $\vec n(z)$ and $\vec n(w)$.

For three unit spinors $z,v,w$, the pairwise overlaps $\braket{z}{v}$, $\braket{v}{w}$, $\braket{w}{z}$ are sometimes called Plücker coordinates \cite{Freidel:2010aq, Freidel:2010tt, Borja:2010rc}: they are invariant under a common $\SU(2)$ rotation of the three spinors, and together with the unit norms they encode the same information as the normals $\vec n(z), \vec n(v), \vec n(w)$ up to that overall rotation. Writing the generator \eqref{eq:app-sp-resolution} as $N(z)=\Id-2\ket{z}\bra{z}$ using the resolution of the identity, and using the trace identity $\Tr\big[(\vec a\cdot\vec\sigma)(\vec b\cdot\vec\sigma)(\vec c\cdot\vec\sigma)\big]=2i\,\vec a\cdot(\vec b\times\vec c)$ for real vectors, the triple product of the three normals is
\begin{equation}
    \vec n(z)\cdot\big(\vec n(v)\times\vec n(w)\big)
    = \frac{1}{2i}\Tr\big(N(z)N(v)N(w)\big)
    = -4\,\Im\big(\braket{z}{v}\braket{v}{w}\braket{w}{z}\big) \, ,
    \label{eq:app-sp-triple-product}
\end{equation}
where, expanding $N(z)N(v)N(w)$ into projectors, the terms with $\Id$ or a single or double projector are real and drop out of the imaginary part, since the left-hand side is manifestly real. Applied to the four area-weighted normals of a closed tetrahedron, \eqref{eq:app-sp-triple-product} expresses its volume directly in terms of spinor overlaps, without ever reconstructing the normals themselves, consistent with the Minkowski reconstruction used in Section~\ref{sec:geometric-constraint-surface}.

A single unit spinor also determines a real unit vector orthogonal to $\vec n(z)$, through
\begin{equation}
    \sbra{z}\vec\sigma\ket{z} = i\vec F(z) + \vec n(z)\times\vec F(z) \, , \qquad \vec n(z)\cdot\vec F(z)=0 \, , \quad |\vec F(z)|=1 \, ,
    \label{eq:app-sp-frame}
\end{equation}
which fixes a reference direction, a framing, in the plane orthogonal to $\vec n(z)$: this is how each spinor determines the framed plane of Section~\ref{sec:setup}. Under the phase change $\ket{z}\to e^{i\alpha}\ket{z}$, the normal $\vec n(z)$ is invariant while $\sbra{z}\vec\sigma\ket{z} \to e^{2i\alpha}\sbra{z}\vec\sigma\ket{z}$, so the frame vector rotates rigidly about $\vec n(z)$ by twice the spinor phase,
\begin{equation}
    \vec F(z) \to R_{\vec n(z)}(2\alpha)\, \vec F(z) \, ,
    \label{eq:app-sp-frame-rotation}
\end{equation}
with $R_{\vec n}(\phi)$ the rotation by angle $\phi$ about $\vec n$: as expected for the $\SU(2)$ double cover, the spinor phase is half the rotation angle of its frame.

As an application, we check directly that the wedge holonomy \eqref{eq:wedge-param} is $\SL(2,\C)$-valued. Since $\braket{z}{z}=1$ and $\sbraket{z}{z}=0$, the pair $(\ket{z},\sket{z})$ is an orthonormal basis of $\C^2$ with $\det(\ket{z}\ \sket{z})=|z^0|^2+|z^1|^2=1$, and likewise $(\sket{\tilde z}, \ket{\tilde z})$ is an orthonormal basis of the target space, with the opposite ordering contributing a compensating sign, $\det(\sket{\tilde z}\ \ket{\tilde z}) = -1$. Using \eqref{eq:app-sp-brackets}, $g_{vf}$ of \eqref{eq:wedge-param} acts on the source basis as
\begin{equation}
    g_{vf}\ket{z_{vf}} = e^{\omega_{vf}/2}\sket{\tilde z_{vf}} \, , \qquad
    g_{vf}\sket{z_{vf}} = -e^{-\omega_{vf}/2}\ket{\tilde z_{vf}} \, ,
    \label{eq:app-sp-unimodular}
\end{equation}
so that $\det g_{vf}$ is the product of the two eigenvalues times the ratio of the two basis determinants, $e^{\omega_{vf}/2}\cdot\big({-}e^{-\omega_{vf}/2}\big)\cdot({-}1) = 1$: the wedge holonomy is automatically unimodular for any choice of unit spinors and complex angle, with no further condition needed.

\section{The EPRL-FK amplitude in wedge variables}
\label{app:vertex-amplitude-details}
To write the transition amplitude \eqref{eq:transition-amplitude} in an exponential form suitable for saddle-point analysis, we use the coherent-state representation of the $\SL(2,\C)$ unitary irreducible representations and introduce auxiliary spinors. The first kind is the \emph{dummy spinor} $\ket{w_{vf}}$, one for each wedge. It implements unitarity of the representation through a resolution of the identity and the corresponding norm weights \cite{Ruhl:1970wnq,Speziale:2016axj}. The second kind is the \emph{external spinor} $\ket{\zeta_{ef}}$, one for each face on each edge. Each wedge $vf$ is bounded by two edges, $e$ and $e'$, and therefore by two such external spinors. The corresponding auxiliary exponent is
\begin{equation}
\label{eq:S-zeta-w}
S^{\zeta w}_\Delta=\sum_{v,f}S^{\zeta w}_{vf}\,,
\qquad
S^{\zeta w}_{vf}
=2j_f\log\!\left(
\frac{\braket{\zeta_{e'f}}{w_{vf}}\bra{w_{vf}}g_{vf}\ket{\zeta_{ef}}}
{\|w_{vf}\|^{1+i\gamma}\|g_{vf}^{\dagger}w_{vf}\|^{1-i\gamma}}
\right)\, .
\end{equation}
Equivalently, each wedge contributes
\begin{equation}
    D^{(\gamma j_{f}, j_{f})}(g_{vf}) = \int \d \mu(w_{vf})\, e^{S^{\zeta w}_{vf}} \, ,
\end{equation}
where $\mu(w_{vf})$ is the measure factor for the $\mathbb{CP}^1$ integration, including the normalization inherited from the Wigner matrices (we refer to the appendices of \cite{Dona:2019dkf} for more details). The path integral contains integrations over both kinds of auxiliary $\mathbb{CP}^1$ spinors: the dummy spinors $w_{vf}$ and the external spinors $\zeta_{ef}$ (plus all the wedge holonomies). To perform the semiclassical analysis of the paper these auxiliary spinors are necessary, but they are not part of the fundamental data of the model. We integrate them out at the level of the saddle-point equations and express the action only in terms of wedge holonomies and face spins.

\medskip

\begin{figure}[H]
\centering
\includegraphics[scale=1]{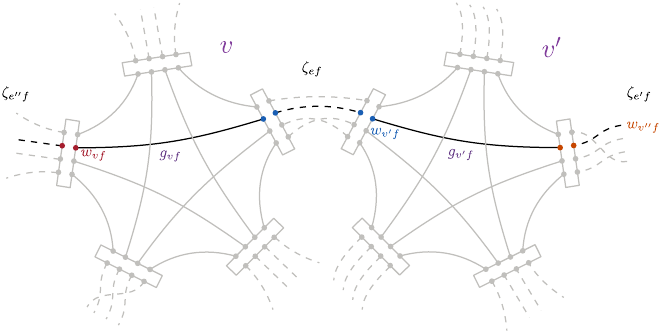}
\caption{\label{fig:two-wedges} Gluing of two adjacent wedges, $vf$ and $v'f$, along the common edge $e$ of the bulk face $f$. Before the external spinor is integrated out, both wedge amplitudes depend on the same spinor $\ket{\zeta_{ef}}$ at $e$. Integrating over this spinor contracts the two wedge contributions and replaces it by the neighboring dummy spinors.}
\end{figure}

We first integrate out the external spinors. Consider an edge $e$ shared by two wedges $vf$ and $v'f$ (see Figure~\ref{fig:two-wedges}). The contributions of both wedges contain the external spinor $\ket{\zeta_{ef}}$, and they contribute to the simplicial complex action as:
\begin{equation}
S^{\zeta w}_{vf}+S^{\zeta w}_{v'f}
= 2 j_{f} \log\!\left(
    \frac{
        \braket{\zeta_{e''f}}{w_{vf}}
        \bra{w_{vf}} g_{vf} \ket{\zeta_{ef}}
    }{
        \|w_{vf}\|^{1+i\gamma}
        \|g_{vf}^{\dagger} w_{vf}\|^{1-i\gamma}
    }
    \right)
    + 2 j_{f} \log\!\left(
    \frac{
        \braket{\zeta_{ef}}{w_{v'f}}
        \bra{w_{v'f}} g_{v'f} \ket{\zeta_{e'f}}
    }{
        \|w_{v'f}\|^{1+i\gamma}
        \|g_{v'f}^{\dagger} w_{v'f}\|^{1-i\gamma}
    }
    \right) \, .
\end{equation}
Integrating out $\ket{\zeta_{ef}}$ contracts the factor in the first logarithm with the dummy spinor of the second wedge, tying the two wedges together:
\begin{equation}
\left.S^{\zeta w}_{vf}+S^{\zeta w}_{v'f}\right|_{\zeta_{ef}}
= 2 j_{f} \log\!\left(
    \frac{
        \braket{\zeta_{e''f}}{w_{vf}}
        \bra{w_{vf}} g_{vf} \ket{w_{v'f}}
    }{
        \|w_{vf}\|^{1+i\gamma}
        \|g_{vf}^{\dagger} w_{vf}\|^{1-i\gamma}
    }
    \right)
    + 2 j_{f} \log\!\left(
    \frac{
        \bra{w_{v'f}} g_{v'f} \ket{\zeta_{e'f}}
    }{
        \|w_{v'f}\|^{1+i\gamma}
        \|g_{v'f}^{\dagger} w_{v'f}\|^{1-i\gamma}
    }
    \right) \, .
\end{equation}

Repeating this procedure at every edge of the two-complex removes all external spinors and leaves an action depending only on the wedge holonomies, face spins, and dummy spinors:
\begin{equation}
    S^{w}_{vf} \equiv \left.S^{\zeta w}_{vf}\right|_{\zeta} = 2 j_{f} \log\!\left( \frac{\bra{w_{vf}} g_{vf} \ket{w_{v'f}}}{\|w_{vf}\|^{1+i\gamma} \|g_{vf}^{\dagger} w_{vf}\|^{1-i\gamma}} \right) \, ,
\end{equation}
where $S^{w}_{vf}$ denotes the reduced wedge exponent after all external spinors have been integrated out. The matrix element of $g_{vf}$ is now taken between the dummy spinors of the wedge itself, as target, and of the previous wedge on the face, as source (see Figure~\ref{fig:two-wedges}).

\medskip

Finally, we extremize the action with respect to the dummy spinors $\ket{w_{v'f}}$ and require the dominance condition. In the full face sum, the first norm in the denominator can be cyclically relabelled,
\begin{equation}
\label{eq:dominance}
S^{w}_\Delta \equiv \sum_{v,f} S^{w}_{vf}
= \sum_{v,f} 2 j_{f} \log\!\left( \frac{\bra{w_{vf}} g_{vf} \ket{w_{v'f}}}{\|w_{vf}\|^{1+i\gamma} \|g_{vf}^{\dagger} w_{vf}\|^{1-i\gamma}} \right) \, , \qquad \text{and} \qquad \Re S^{w}_\Delta = 0 \, .
\end{equation}
where $v'$ is the vertex before $v$ in the face $f$ and $v''$ is the vertex before $v'$ in the face $f$. Since $w_{v'f}$ appears exclusively in $S^{w}_{vf}$ and $S^{w}_{v'f}$, the variation of $S^{w}_\Delta$ reduces to the variation of these two terms. The resulting equations are
\begin{align}
\label{eq:alignment}
\frac{\delta (S^{w}_{vf}+S^{w}_{v'f})}{\delta \bra{w_{v'f}}} &= 2j_f \left[ \frac{g_{v'f}\ket{w_{v''f}}}{\bra{w_{v'f}} g_{v'f} \ket{w_{v''f}}} - \frac{1+i\gamma}{2}\frac{\ket{w_{v'f}}}{\braket{w_{v'f}}{w_{v'f}}} - \frac{1-i\gamma}{2}\frac{g_{v'f} g_{v'f}^{\dagger}\ket{w_{v'f}}}{\bra{w_{v'f}} g_{v'f} g_{v'f}^{\dagger} \ket{w_{v'f}}} \right] = 0 \, ,\\
\frac{\delta (S^{w}_{vf}+S^{w}_{v'f})}{\delta \ket{w_{v'f}}} &= 2j_f \left[ \frac{\bra{w_{vf}} g_{vf}}{\bra{w_{vf}} g_{vf} \ket{w_{v'f}}} - \frac{1+i\gamma}{2}\frac{\bra{w_{v'f}}}{\braket{w_{v'f}}{w_{v'f}}} - \frac{1-i\gamma}{2}\frac{\bra{w_{v'f}} g_{v'f} g_{v'f}^{\dagger}}{\bra{w_{v'f}} g_{v'f} g_{v'f}^{\dagger} \ket{w_{v'f}}} \right] = 0 \, .
\end{align}
The real part of the action is
\begin{equation}
\Re S^{w}_\Delta = \sum_{v,f} 2 j_{f} \log\!\left( \frac{|\bra{w_{vf}} g_{vf} \ket{w_{v'f}}|}{\|w_{v'f}\| \, \|g_{vf}^{\dagger} w_{vf}\|} \right) \, ,
\end{equation}
where we have used the cyclically relabelled form of \eqref{eq:dominance}. Thus no additional factor of $\|w_{vf}\|$ appears in this expression: the denominator is precisely the product of the norms of the two spinors entering the scalar product, $g_{vf}^{\dagger}w_{vf}$ and $w_{v'f}$. By the Cauchy--Schwarz inequality this quantity is non-positive, and it vanishes if and only if every term in the sum vanishes, i.e. if
\begin{equation}
\label{eq:eigen-gdagger}
g_{vf}^\dagger \ket{w_{vf}} = \kappa_{vf} \, \ket{w_{v'f}} \, , \qquad 
\end{equation}
with $\kappa_{vf}$ a complex number. The companion relation for $g_{vf}$ follows directly from the bra-variation of $S^{w}_\Delta$,
\begin{equation}
\label{eq:alignment-vf}
\frac{\delta S^{w}_\Delta}{\delta \bra{w_{vf}}} = 2j_f \left[ \frac{g_{vf}\ket{w_{v'f}}}{\bra{w_{vf}} g_{vf} \ket{w_{v'f}}} - \frac{1+i\gamma}{2}\frac{\ket{w_{vf}}}{\|w_{vf}\|^2} - \frac{1-i\gamma}{2}\frac{g_{vf} g_{vf}^{\dagger}\ket{w_{vf}}}{\|g_{vf}^{\dagger} w_{vf}\|^2} \right] = 0 \, ,
\end{equation}
once \eqref{eq:eigen-gdagger} is substituted in. 
\begin{equation}
\frac{g_{vf}\ket{w_{v'f}}}{\overline{\kappa_{vf}}\,\|w_{v'f}\|^2} - \frac{1+i\gamma}{2}\frac{\ket{w_{vf}}}{\|w_{vf}\|^2} - \frac{1-i\gamma}{2}\kappa_{vf} \frac{g_{vf} \ket{w_{v'f}}}{|\kappa_{vf}|^2 \|w_{v'f}\|^2} = 0 \, ,
\end{equation}
multiplying by $\overline{\kappa_{vf}}$ and factorizing gives
\begin{equation}
\frac{g_{vf}\ket{w_{v'f}}}{\|w_{v'f}\|^2}  = \overline{\kappa_{vf}}\frac{\ket{w_{vf}}}{\|w_{vf}\|^2} \, ,
\end{equation} 
or equivalently
\begin{equation}
g_{vf}\ket{w_{v'f}} = \overline{\kappa_{vf}}\,\frac{\|w_{v'f}\|^2}{\|w_{vf}\|^2}\ket{w_{vf}} \, , \qquad
g_{vf}^\dagger \ket{w_{vf}} = \kappa_{vf} \, \ket{w_{v'f}} \, .
\end{equation}
These equations determine the dummy spinors as elements of $\mathbb{CP}^{1}$. In other words, they fix $\ket{w_{v'f}}$ and $\ket{w_{vf}}$ only up to non-zero complex rescalings. Choosing unit representatives in each projective class, the wedge parametrization \eqref{eq:wedge-param} gives, on the branch used in the main text,
\begin{equation}
\ket{w_{v'f}} = \ket{z_{vf}} \, , \qquad
\ket{w_{vf}} = \sket{\tilde z_{vf}} \, ,
\end{equation}
with $\kappa_{vf}=e^{\overline{\omega}_{vf}/2}$. The other branch corresponds to the unit representatives $\ket{w_{v'f}}=\sket{z_{vf}}$ and $\ket{w_{vf}}=\ket{\tilde z_{vf}}$.

The same argument applied to the next wedge $v'f$ gives another equation for the same dummy spinor,
\begin{equation}
\ket{w_{v'f}} = \sket{\tilde z_{v'f}} \, .
\end{equation}
Thus integrating out the dummy spinors also imposes the compatibility condition between consecutive wedge holonomies,
\begin{equation}
\ket{z_{vf}} = \sket{\tilde z_{v'f}} \, ,
\end{equation}
after fixing the residual phase of the $\mathbb{CP}^{1}$ representatives. This is the spinor identification used in the main text.

Using the wedge parametrization \eqref{eq:wedge-param} and the same unit representatives, the matrix element in the wedge exponent evaluates to
\begin{equation}
\bra{w_{vf}} g_{vf} \ket{w_{v'f}}
= \bra{w_{vf}} \left(  e^{\omega_{vf}/2} \sket{\tilde z_{vf}}\bra{z_{vf}} - e^{-\omega_{vf}/2} \ket{\tilde z_{vf}}\sbra{z_{vf}}  \right) \ket{w_{v'f}}
= e^{\omega_{vf}/2} \, ,
\end{equation}
with norms
\begin{equation}
    \| w_{vf} \|^2 = 1 \, , \qquad
    \| g_{vf}^{\dagger} w_{vf} \|^2 = \bra{w_{vf}} g_{vf} g_{vf}^{\dagger} \ket{w_{vf}} = e^{\Re\omega_{vf}} \, .
\end{equation}
The logarithm in the wedge exponent therefore evaluates to a phase,
\begin{equation}
    \log\!\left( \frac{\bra{w_{vf}} g_{vf} \ket{w_{v'f}}}{\|w_{vf}\|^{1+i\gamma} \|g_{vf}^{\dagger} w_{vf}\|^{1-i\gamma}} \right)
    =
    \log\!\left(\frac{e^{\omega_{vf}/2}}{e^{(1-i\gamma)\Re\omega_{vf}/2}}\right)
    =
    \frac{i}{2}\left(\gamma\,\Re\omega_{vf}+\Im\omega_{vf}\right) \, .
\end{equation}
Multiplying by the prefactor $2j_f$ and summing over wedges gives
\begin{equation}
S^{\EPRLwedge} = i\sum_{f} j_{f} \sum_{v\subset f}\left( \gamma \, \Re\,\omega_{vf} + \Im\,\omega_{vf} \right) \, .
\end{equation}

\section{Detailed derivation of the equations of motion of the EPRL-FK wedge action}
\label{app:eom-details}

To study the semiclassical limit of the theory, we compute the equations of motion of the action~\eqref{eq:actionEPRL}. Varying with respect to the Lagrange multipliers $\mu_{vabc}$ returns the local flatness constraint, \eqref{eq:local-flatness}, which fixes the complex angles $\omega_{vf}$ in terms of the spinors, \eqref{eq:omega-flatness}, and imposes shape matching as discussed above. Varying with respect to the complex angle $\omega_{vf}$ gives
\begin{equation}
    \label{eq:app-omega-var}
    \frac{\delta S^{\text{EPRL-FK}}}{\delta \omega_{vf}} = i\left[
    \frac{j_{f}}{2} \left( \gamma - i \right) + \sum_{vabc\supset vf} \, \Tr\left( \mu_{vabc}\, g_{v f_{ac}}\, g_{v f_{cb}}\, \frac{\delta g_{v f}}{\delta \omega_{vf}} \right)
    \right] = 0 \, .
\end{equation}
where, for simplicity, we have assumed that the wedge $vf$ is the first in the cycle $abc$. The derivative of the wedge holonomy with respect to the complex angle is
\begin{equation}
    \frac{\delta g_{v f}}{\delta \omega_{vf}} = \frac{1}{2} \left( e^{\omega_{vf}/2} \sket{\tilde z_{vf}}\bra{z_{vf}} + e^{-\omega_{vf}/2} \ket{\tilde z_{vf}}\sbra{z_{vf}} \right) \, .
\end{equation}
Its contraction with the inverse wedge holonomy is
\begin{equation}
    g_{v f}^{-1} \frac{\delta g_{v f}}{\delta \omega_{vf}} = \frac{1}{2} \left( \ket{z_{vf}}\bra{z_{vf}} - \sket{z_{vf}}\sbra{z_{vf}} \right) = - \frac{1}{2}N_{vf} \, ,
\end{equation}
where $ N_{vf}$ is the $\SL(2,\C)$-algebra element associated to the spinor $\ket{z_{vf}}$ of the wedge $vf$. In terms of the normal $\vec{n}_{vf}$ we can write $N_{vf} = \vec{n}_{vf}\cdot\vec{\sigma}$ (see Appendix~\ref{app:spinors}), where $\vec{\sigma}$ are the Pauli matrices.
On shell of local flatness, we have $g_{v f_{ac}}\, g_{v f_{cb}}\, g_{v f} = \Id$, so that $g_{v f_{ac}}\, g_{v f_{cb}} = g_{v f}^{-1}$. This allows us to rewrite the second term of \eqref{eq:app-omega-var} as
\begin{equation}
    \frac{\delta S^{\text{EPRL-FK}}}{\delta \omega_{vf}} = i\left[
    \frac{j_{f}}{2} \left( \gamma - i \right) - \frac{1}{2} \sum_{vabc\supset vf} \, \Tr\left( \mu_{vabc}\, N_{vf} \right)
    \right] = 0\, .
\end{equation}
Dropping the common factor of $i$, dropping the factor $2$, and moving the second term to the right-hand side, we can write this equation as
\begin{equation}
    \label{eq:app-omega-var-trace}
    j_{f} \left( \gamma - i \right) =  \, \Tr\left( \left( \sum_{vabc\supset vf} \mu_{vabc} \right) \, N_{vf} \right) \, .
\end{equation}
If we package the Lagrange multipliers $\mu_{vabc}$ into a single object per wedge,
\begin{equation}
\mu_{vf} \equiv \sum_{vabc\supset vf} \mu_{vabc} \, ,
\end{equation}
then \eqref{eq:app-omega-var-trace} simplifies to
\begin{equation}
    j_{f} \left( \gamma - i \right) =  \, \Tr\left( \mu_{vf} \, N_{vf} \right) \, .
\end{equation}
This fixes the component of the Lagrange multiplier $\mu_{vf}$ along the normal $N_{vf}$ in terms of the spin $j_f$. The other two components of $\mu_{vf}$, orthogonal to $N_{vf}$, remain undetermined. Without the assumption that the wedge $vf$ is the first in the cycle $abc$, $\mu_{vf}$ is the sum of three Lagrange multipliers parallel transported to the source of the wedge $vf$.

We now vary with respect to the normalized spinor $\ket{z_{ef}}$ shared by two consecutive wedges of the same face: it is the source spinor of $g_{vf}$ and the target spinor of $g_{v'f}$. A tangent variation $\delta\ket{z_{ef}}=X\ket{z_{ef}}$, with $X$ an $\SU(2)$ algebra element, induces, at fixed complex angles,
\begin{equation}
    \delta g_{vf}=-g_{vf}X \, ,     \qquad      \delta g_{v'f}=X g_{v'f}\, .
\end{equation}
The wedge part of the action has no explicit spinor dependence, so only the local-flatness terms vary,
\begin{align}
    \delta S^{\text{EPRL-FK}} &= i\Tr\left(\mu_{vf}g_{vf}^{-1}\delta g_{vf} \right) + i\Tr\left( \mu_{v'f}g_{v'f}^{-1}\delta g_{v'f} \right) \nonumber \\
    &= i\Tr\left[\left(-\mu_{vf} + \mathrm{Ad}_{g_{v'f}}\mu_{v'f}\right)X \right] \, .
\end{align}
Since $X$ is arbitrary, stationarity requires
\begin{equation}
\label{eq:app-muparalleltransport}
\mu_{v'f} = \mathrm{Ad}_{g_{v'f}^{-1}}\mu_{vf} \, :
\end{equation}
the Lagrange multipliers of two consecutive wedges are related by parallel transport.

This is consistent with the geometric content of local flatness. Writing $N_{vf}$ and $\tilde N_{vf}$ for the generators of the source and target normals of $g_{vf}$, the wedge parametrization gives $\mathrm{Ad}_{g_{v'f}}N_{v'f} = -\tilde N_{v'f}$. Dualizing a target spinor into a source spinor reverses the sign of its generator, so $\tilde N_{v'f} = -N_{vf}$ (recall $\ket{z_{ef}} = \sket{\tilde z_{v'f}} = \ket{z_{vf}}$, so that $N_{vf}=N_{ef}$). Hence $\mathrm{Ad}_{g_{v'f}}N_{v'f} = N_{ef}$: the normals obey exactly the same transport law as the multipliers, \eqref{eq:app-muparalleltransport}. The $\omega$-equations then fix the multipliers' component along the normal in terms of the spins,
\begin{equation}
    \mu_{vf} = \frac{1}{2} j_{f} \left( \gamma - i \right)  N_{ef}  + \mu_{vf}^{\perp} \, ,
\end{equation}
where the $1/2$ compensates $\Tr(N_{ef}^2)=2$, leaving only the transverse part $\mu_{vf}^{\perp}$ undetermined. \eqref{eq:app-muparalleltransport} then gives
\begin{equation}
\mu_{v'f}^{\perp} = \mathrm{Ad}_{g_{v'f}^{-1}} \mu_{vf}^{\perp} \, ,
\end{equation}
so that, iterating around the whole face back to the original wedge,
\begin{equation}
\mu_{vf}^{\perp} = \mathrm{Ad}_{g_{f}^{-1}} \mu_{vf}^{\perp} \, ,
\end{equation}
where $g_f$ is the holonomy around the face. The face holonomy is a $4$-screw stabilizing the bivector of the triangle, so the only solution is $\mu_{vf}^{\perp}=0$. The wedge Lagrange multipliers are therefore completely fixed,
\begin{equation}
    \label{eq:app-mu-fixed}
    \mu_{vf} = \frac{1}{2} j_{f} \left( \gamma - i \right)  N_{ef} \, .
\end{equation}
The multipliers $\mu_{vf}$ are enough to determine all the $\mu_{vabc}$ if we choose a delta function for each fundamental cycle. Fix a vertex $v$ and choose a maximal tree $T$ in its graph of wedge holonomies. Every wedge $vc\notin T$, called a chord, determines a unique fundamental cycle $C_{vabc}$, obtained by joining $vc$ with the unique path in $T$ connecting its endpoints. By construction, $vc$ belongs to no other fundamental cycle. Choosing the source of $vc$ as the base point of $C_{vabc}$, we may order the cycle holonomy to have a chord holonomy $g_{vc}$ as the first term. Since $g_{vc}$ appears in no other fundamental cycle, its effective wedge multiplier $\mu_{vc}$ coincides with $\mu_{vabc}$. The wedge equation then determines the cycle multiplier directly:
\begin{equation}
\label{eq:app-chord-mu}
\mu_{vabc} \equiv \mu_{vc}=
\frac{j_{c}}{2}(\gamma - i)N_{vc} \, .
\end{equation}
Thus, the equations associated with the chords determine all the independent cycle multipliers.

Once the cycle multipliers have been determined, the equations associated with the links of the maximal tree become compatibility conditions. We spell this out explicitly for a vertex with edges $1,2,3,4,5$ and the maximal tree $T$ rooted at edge $1$: tree links $vf_{21}, vf_{31}, vf_{41}, vf_{51}$, each oriented with $1$ as source, and chords $vf_{32}, vf_{42}, vf_{52}, vf_{43}, vf_{53}, vf_{54}$, each oriented with the lower-numbered edge as source.

Consider edge $2$. It is shared by the tree link $vf_{21}$ and by the three chords $vf_{32}, vf_{42}, vf_{52}$, each belonging to a single fundamental cycle, respectively $(1,2,3)$, $(1,2,4)$, and $(1,2,5)$, based, as above, at the chord's own source, edge $2$. By \eqref{eq:app-chord-mu} these three cycle multipliers are already known,
\begin{equation}
\begin{aligned}
\mu_{v231}=\mu_{vf_{32}}&=-i\tfrac12 j_{23}(1+i\gamma)N_{vf_{32}}\, , \\
\mu_{v241}=\mu_{vf_{42}}&=-i\tfrac12 j_{24}(1+i\gamma)N_{vf_{42}}\, , \\
\mu_{v251}=\mu_{vf_{52}}&=-i\tfrac12 j_{25}(1+i\gamma)N_{vf_{52}}\, ,
\end{aligned}
\end{equation}
each normal evaluated at edge $2$.

The same three cycles also contain the tree link $vf_{21}$, whose own source is edge $1$ rather than $2$. Rewriting each cyclic product $g_{vf_{21}}g_{vf_{1y}}g_{vf_{y2}}=\Id$ ($y=3,4,5$, from \eqref{eq:local-flatness}) to start instead from $vf_{21}$, cyclicity of the trace gives that cycle's contribution to $\mu_{vf_{21}}$ as $\mathrm{Ad}_{g_{vf_{21}}^{-1}}\mu_{vf_{y2}}$. Summing the three cycles and comparing with the general wedge equation \eqref{eq:app-mu-fixed} applied directly to $vf_{21}$,
\begin{equation}
-i\tfrac12 j_{12}(1+i\gamma)N_{vf_{21}} = \mu_{vf_{21}} = \mathrm{Ad}_{g_{vf_{21}}^{-1}}\Big(\mu_{vf_{32}}+\mu_{vf_{42}}+\mu_{vf_{52}}\Big) \, .
\end{equation}
Cancelling $\tfrac12(1+i\gamma)$ and applying $\mathrm{Ad}_{g_{vf_{21}}}$ to both sides,
\begin{equation}
j_{12}\, \mathrm{Ad}_{g_{vf_{21}}}N_{vf_{21}} = j_{23}N_{vf_{32}}+j_{24}N_{vf_{42}}+j_{25}N_{vf_{52}} \, .
\end{equation}
The left-hand side is exactly the source/target relation used above, $\mathrm{Ad}_{g_{vf_{21}}}N_{vf_{21}}=-\tilde N_{vf_{21}}=-N_{vf_{12}}$: minus the same wedge's normal, now evaluated at edge $2$ instead of edge $1$. Hence
\begin{equation}
j_{12}N_{vf_{12}} + j_{23}N_{vf_{32}}+j_{24}N_{vf_{42}}+j_{25}N_{vf_{52}} = 0 \, ,
\end{equation}
which is precisely the closure equation \eqref{eq:closure-spinorial} for the tetrahedron at edge $2$: once every wedge touching $2$ is expressed with $2$ as source, all four enter with the same sign.

The identical mechanism, summing the known chord multipliers around a leaf, transporting the tree link into the leaf's own frame via cyclicity and \eqref{eq:local-flatness}, and closing with $\tilde N_{v'f}=-N_{vf}$, gives the closure equation at each of the other three leaves, $3$, $4$, and $5$; we have checked this explicitly for edge $5$ as well, where the bookkeeping differs slightly (there, none of the three chords touching $5$ is already oriented with $5$ as source, so the sign cancellation is produced by the transport step itself rather than by a final dualization), but the conclusion is the same clean closure equation. Closure at the root, edge $1$, is not obtained independently this way; it follows instead from the four leaf equations together with the vertex's overall gauge symmetry (Section~\ref{sec:symmetries}), which makes one of the five closure conditions redundant.

\section{Geometric reconstruction lemmas}
\label{app:geometric-reconstruction-lemmas}

\subsection{Angle sum for the reconstructed triangles}
\label{app:triangle-angle-sum-proof}
Let $f,g,h,k$ be the four faces of a geometrically closed, nondegenerate tetrahedron, and suppress the common tetrahedron label on their spinors. For each face $a\neq f$, the complex number $\sbrasket{z_f}{z_a}\sbraket{z_f}{z_a}$ has a phase that gives the oriented direction of the edge $f\cap a$ in the plane of $f$. Choose the cyclic ordering of $g,h,k$ induced by the orientation of $f$. The external two-dimensional angles are defined by
\begin{equation}
    \alpha^f_{gh}
    =
    \arg\left(\frac{\sbrasket{z_f}{z_g}\sbraket{z_f}{z_g}}{\sbrasket{z_f}{z_h}\sbraket{z_f}{z_h}}\right)\, ,
    \qquad
    \alpha^f_{hk}
    =
    \arg\left(\frac{\sbrasket{z_f}{z_h}\sbraket{z_f}{z_h}}{\sbrasket{z_f}{z_k}\sbraket{z_f}{z_k}}\right)\, ,
    \qquad
    \alpha^f_{kg}
    =
    \arg\left(\frac{\sbrasket{z_f}{z_k}\sbraket{z_f}{z_k}}{\sbrasket{z_f}{z_g}\sbraket{z_f}{z_g}}\right)\, ,
    \label{eq:app-external-angle-phases}
\end{equation}
where the branches are chosen so that $0<\alpha^f_{gh},\alpha^f_{hk},\alpha^f_{kg}<\pi$. This is possible because $f$ is a nondegenerate convex Euclidean triangle.

Using \eqref{eq:app-external-angle-phases} and additivity of the argument modulo $2\pi$, the sum of the three angles is
\begin{align}
    \alpha^f_{gh}+\alpha^f_{hk}+\alpha^f_{kg}
    &=
    \arg\left(\frac{\sbrasket{z_f}{z_g}\sbraket{z_f}{z_g}}{\sbrasket{z_f}{z_h}\sbraket{z_f}{z_h}}\right)
    +\arg\left(\frac{\sbrasket{z_f}{z_h}\sbraket{z_f}{z_h}}{\sbrasket{z_f}{z_k}\sbraket{z_f}{z_k}}\right)
    +\arg\left(\frac{\sbrasket{z_f}{z_k}\sbraket{z_f}{z_k}}{\sbrasket{z_f}{z_g}\sbraket{z_f}{z_g}}\right)
    \notag\\
    &=
    \arg(1)
    =0
    \pmod{2\pi}\, .
\end{align}
Since all three external angles lie strictly between $0$ and $\pi$, their sum lies strictly between $0$ and $3\pi$, so the only allowed multiple of $2\pi$ is
\begin{equation}
    \alpha^f_{gh}+\alpha^f_{hk}+\alpha^f_{kg}=2\pi
    \, .
    \label{eq:app-external-triangle-angle-sum}
\end{equation}

\subsection{Independence of the reconstructed edge scalar}
\label{app:edge-scalar-reconstruction-independence}
We prove directly in spinor variables that the reconstructed edge scalar is independent of the triangle used in its reconstruction. We first compare the two triangular faces of a single tetrahedron containing the edge, and then propagate the equality through its connected star.

Let $f,g,h,k$ denote the four triangular faces of a tetrahedron, suppressing the common tetrahedron label on their unit spinors. Introduce the rank-one projectors (the special case $u=v=z_a$ of the Fierz identity of Appendix~\ref{app:spinors})
\begin{equation}
    P_a\equiv\ket{z_a}\bra{z_a}\, ,
    \qquad a\in\{f,g,h,k\}\, .
    \label{eq:app-spinor-projectors}
\end{equation}
For unit spinors, geometric closure is equivalent to
\begin{equation}
    \sum_{a=f,g,h,k}j_aP_a
    =
    \frac12\left(\sum_{a=f,g,h,k}j_a\right)\Id\, .
    \label{eq:app-projector-closure}
\end{equation}

For a reference face $r$ and two other faces $a,b$, define
\begin{equation}
    \chi_a^r
    \equiv
    \sbrasket{z_r}{z_a}\,
    \sbraket{z_r}{z_a}\, ,
    \qquad
    \Delta_{ab}^r
    \equiv
    \Im\!\left(\chi_a^r\overline{\chi_b^r}\right)\, .
    \label{eq:app-chi-delta}
\end{equation}
A direct two-spinor calculation gives
\begin{equation}
    \Delta_{ab}^r
    =
    -\Im\Tr(P_rP_aP_b)
    =
    -\Im\!\left(
        \braket{z_r}{z_a}
        \braket{z_a}{z_b}
        \braket{z_b}{z_r}
    \right)\, .
    \label{eq:app-delta-projectors}
\end{equation}
The explicit spinorial form of the three sines in
\eqref{eq:reconstructed-edge-scalar} therefore gives
\begin{equation}
    s_{f\cap k}^{(f)}
    =
    2j_f|\chi_k^f|^2
    \frac{|\Delta_{gh}^f|}
    {|\Delta_{gk}^f|\,|\Delta_{hk}^f|}\, .
    \label{eq:app-edge-scalar-spinorial}
\end{equation}

Multiply \eqref{eq:app-projector-closure} by $P_fP_g$, take the trace and then its imaginary part. The right-hand side is real, as are the terms proportional to $j_f$ and $j_g$. Cyclicity of the trace and \eqref{eq:app-delta-projectors} give
\begin{equation}
    j_h\Delta_{gh}^f+j_k\Delta_{gk}^f=0\, .
    \label{eq:app-delta-closure-one}
\end{equation}
Repeating the calculation with $P_fP_h$ gives
\begin{equation}
    j_g\Delta_{gh}^f-j_k\Delta_{hk}^f=0\, .
    \label{eq:app-delta-closure-two}
\end{equation}
Consequently,
\begin{equation}
    |\Delta_{gk}^f|
    =
    \frac{j_h}{j_k}|\Delta_{gh}^f|\, ,
    \qquad
    |\Delta_{hk}^f|
    =
    \frac{j_g}{j_k}|\Delta_{gh}^f|\, ,
    \label{eq:app-delta-closure-ratios}
\end{equation}
and hence
\begin{equation}
    s_{f\cap k}^{(f)}
    =
    \frac{2j_fj_k^2}{j_gj_h}
    \frac{|\chi_k^f|^2}{|\Delta_{gh}^f|}\, .
    \label{eq:app-edge-scalar-from-f}
\end{equation}

The reconstruction of the same edge from the other incident triangle $k$ is obtained by exchanging $f$ and $k$:
\begin{equation}
    s_{f\cap k}^{(k)}
    =
    \frac{2j_kj_f^2}{j_gj_h}
    \frac{|\chi_f^k|^2}{|\Delta_{gh}^k|}\, .
    \label{eq:app-edge-scalar-from-k}
\end{equation}
The elementary symmetry properties of the spinor contractions imply
\begin{equation}
    |\chi_k^f|^2
    =
    \left|\sbrasket{z_f}{z_k}\right|^2
    \left|\sbraket{z_f}{z_k}\right|^2
    =
    \left|\sbrasket{z_k}{z_f}\right|^2
    \left|\sbraket{z_k}{z_f}\right|^2
    =
    |\chi_f^k|^2\, .
    \label{eq:app-chi-symmetry}
\end{equation}

Finally, multiply \eqref{eq:app-projector-closure} by $P_gP_h$, take the trace and then its imaginary part. This gives
\begin{equation}
    j_f\Delta_{gh}^f+j_k\Delta_{gh}^k=0\, ,
    \qquad\Longrightarrow\qquad
    j_f|\Delta_{gh}^f|
    =
    j_k|\Delta_{gh}^k|\, .
    \label{eq:app-delta-opposite-faces}
\end{equation}
Using \eqref{eq:app-chi-symmetry} and
\eqref{eq:app-delta-opposite-faces} in
\eqref{eq:app-edge-scalar-from-k}, we find
\begin{align}
    s_{f\cap k}^{(k)}
    &=
    \frac{2j_kj_f^2}{j_gj_h}
    \frac{|\chi_k^f|^2}
    {(j_f/j_k)|\Delta_{gh}^f|}
    \nonumber\\
    &=
    \frac{2j_fj_k^2}{j_gj_h}
    \frac{|\chi_k^f|^2}{|\Delta_{gh}^f|}
    =
    s_{f\cap k}^{(f)}\, .
    \label{eq:app-edge-scalar-same-tetrahedron}
\end{align}
Thus geometric closure makes the reconstruction independent of which of the two triangles containing the edge is used inside a fixed tetrahedron.

It remains to compare different tetrahedra and four-simplices. Let two tetrahedra inside the same four-simplex share a triangle $r$. On the geometric branch, local flatness imposes spinorial angle matching, so the corresponding invariants
\begin{equation}
    \frac{|\Delta_{ab}^r|}
    {|\chi_a^r|\,|\chi_b^r|}
    =
    \sin\alpha_{ab}^r
    \label{eq:app-spinorial-angle-matching}
\end{equation}
agree when computed from either tetrahedron. Since the two reconstructions also use the same spin $j_r$, \eqref{eq:reconstructed-edge-scalar} assigns the same scalar to every edge of the shared triangle.

Adjacent four-simplices use the same spinors and spins on their common tetrahedron, so their reconstructions agree as well. If the star of a triangulation edge $p$ is connected, these equalities propagate along a chain of incident tetrahedra and four-simplices. Hence, on the nondegenerate geometric constraint surface,
\begin{equation}
    s_p^{(e,f)}
    =
    s_p^{(e',f')}
    \qquad
    \text{for all $f,f'\supset p$ and $e,e'\supset p$}\, .
    \label{eq:app-global-edge-scalar-independence}
\end{equation}
The reconstructed scalar $s_p$ is therefore well defined.

\section{A particle with conserved angular momentum as a strong constraint}
\label{app:particle-angular-momentum}
Consider a free particle moving in the plane, with action
\begin{equation*}
S_0[x,y] = \int \d t\, \frac{m}{2}\left(\dot x^{2}+\dot y^{2}\right)\, .
\end{equation*}
Its equations of motion are simply
\begin{equation*}
\ddot x = 0\, , \qquad \ddot y = 0\, .
\end{equation*}
The solutions are straight lines with constant velocity, since there are no forces. Rotational invariance also implies conservation of the angular momentum
\begin{equation*}
J = m\left(x\dot y-y\dot x\right)\, ,
\end{equation*}
since, on the equations of motion,
\begin{equation*}
\dot J = m\left(x\ddot y-y\ddot x\right) = 0\, .
\end{equation*}
Thus $\dot J=0$ is a consequence of the equations of motion, but strictly weaker than them.

Consider a deceptively harmless modification: adding $\dot J$ to the Lagrangian with a \emph{constant} coefficient $\alpha$ changes nothing, since
\begin{equation*}
\int \d t\, \alpha \dot J = \alpha J \big|_{t_i}^{t_f}
\end{equation*}
is a pure boundary term. One might expect the same to hold once $\alpha$ is promoted to a time-dependent Lagrange multiplier $\lambda(t)$, since $\lambda \dot J$ still looks like ``adding zero'' whenever $J$ is conserved. This expectation fails, as we now show. Define the extended action
\begin{equation*}
S_{\mathrm{ext}} = S_0 + \int \d t\, \lambda(t)\, \dot J\, .
\end{equation*}
Integrating by parts and dropping the boundary term gives
\begin{equation*}
S_{\mathrm{ext}} = \int \d t\, \left(\frac{m}{2}\left(\dot x^{2}+\dot y^{2}\right) - \dot\lambda\, J\right)\, .
\end{equation*}
Unlike the constant-$\alpha$ case, $\dot\lambda$ now couples dynamically to $J$, and the resulting equations of motion differ genuinely from the free ones. In polar coordinates we can find a simple solution to the extended equations that is not a solution of the original free-particle equations. With
\begin{equation*}
x = r\cos\theta\, , \qquad y = r\sin\theta\, , \qquad J = m r^{2}\dot\theta\, ,
\end{equation*}
the extended Lagrangian reads
\begin{equation*}
\mathcal{L}_{\mathrm{ext}} = \frac{m}{2}\left(\dot r^{2}+r^{2}\dot\theta^{2}\right) - m\dot\lambda\, r^{2}\dot\theta\, .
\end{equation*}
Varying $\lambda$, $\theta$, and $r$ gives, respectively,
\begin{equation*}
\frac{\d}{\d t}\left(m r^{2}\dot\theta\right) = 0\,, \quad
\frac{\d}{\d t}\left[m r^{2}\left(\dot\theta-\dot\lambda\right)\right] = 0\, , \quad
\ddot r = r\dot\theta^{2} - 2\dot\lambda\, r\dot\theta\, .
\end{equation*}
The first equation only enforces angular-momentum conservation, as before. The other two are \emph{not} the free-particle equations: the multiplier acts as an effective force through the terms proportional to $\dot\lambda$. Indeed,
\begin{equation*}
r(t) = R\, , \qquad \theta(t) = \Omega t\, , \qquad \lambda(t) = \frac{\Omega}{2}\,t \, ,
\end{equation*}
solves the extended equations for any constants $R,\Omega$. Its projection onto the original variables,
\begin{equation*}
x(t) = R\cos(\Omega t)\, , \qquad y(t) = R\sin(\Omega t)\, ,
\end{equation*}
describes uniform circular motion, manifestly not a solution of the free particle equations. The extended action therefore admits trajectories absent from the original theory. The two theories agree only on the sector $\dot\lambda=0$, to which the extended dynamics is not confined.

\section{A particle on a circle as a constrained system}
\label{app:particle-circle}
This appendix gives a simple mechanical model of the constrained variational argument used in the main text. We compare two descriptions of the same physical system: an intrinsic particle whose configuration variable is an angle $\theta\in S^1$, and an ambient particle $q=(x,y)\in\R^2$ whose motion is restricted to the circle of radius $R$ by a Lagrange multiplier. The two descriptions have exactly the same physical trajectories once the auxiliary multiplier is projected out.

\medskip

Let $I=[t_i,t_f]$ be a time interval, with variations of the dynamical variables vanishing at $t_i$ and $t_f$. We assume the particle moves in a smooth potential $U$. The circle is parametrized by an angle $\theta$, embedded in the plane as
\begin{equation}
    \label{eq:circle-embedding}
    q(\theta)=R
    \begin{pmatrix}
        \cos\theta\\
        \sin\theta
    \end{pmatrix}\, .
\end{equation}

Let $U_R(\theta) \equiv U(q(\theta))$ be the potential restricted to the circle. The intrinsic action is
\begin{equation}
    \label{eq:circle-intrinsic-action}
    S_{S^1}[\theta]
    =\int_I \d t\,
    \left[
        \frac{mR^2}{2}\dot\theta^2-U_R(\theta)
    \right]\, .
\end{equation}
Taking variations of $\theta$ that vanish at the endpoints gives the intrinsic equation of motion
\begin{equation}
    \label{eq:circle-intrinsic-eom}
    \left(mR^2\ddot\theta+\frac{\d U_R}{\d\theta}\right)\delta\theta=0\, .
\end{equation}

\medskip

We now treat $q$ as an unconstrained variable in $\R^2$, and impose that it stays on the circle through the constraint
\begin{equation}
    \label{eq:circle-constraint}
    C(q) \equiv q^2-R^2=0\, .
\end{equation}
Introducing a Lagrange multiplier $\lambda(t)$ to enforce this constraint, the extended action is
\begin{equation}
    \label{eq:circle-extended-action}
    S_{\mathrm{ext}}[q,\lambda]
    =\int_I\d t\,\left(\frac{m}{2}\dot q^2-U(q) - \frac{\lambda}{2}\bigl(q^2-R^2\bigr)\right)\, .
\end{equation}
Taking variations of $q$ and $\lambda$ that vanish at the endpoints gives the extended equations of motion
\begin{equation}
    \label{eq:circle-constrained-eom}
    \left(m\ddot q+\nabla U(q) - \lambda q\right)\cdot \delta q + \frac{1}{2}\bigl(q^2-R^2\bigr)\delta \lambda=0\, .
\end{equation}
The dynamical variables span a $3$-dimensional space $(q,\lambda)\in\R^2\times\R$, but the physical motion is only $1$-dimensional. One dynamical variable is eliminated by the constraint, and another is eliminated by the Lagrange multiplier $\lambda$, which has no physical content of its own. The physical degree of freedom is the angle $\theta$ along the circle. Setting to zero the coefficient of $\delta \lambda$ recovers the constraint \eqref{eq:circle-constraint}, $q^2=R^2$. 
The variation $\delta q$ is two dimensional. It is useful to decompose it into a tangential direction, tangent to the constraint surface (the circle) and satisfying \eqref{eq:circle-linearized-constraint}, and a radial direction, normal to the constraint surface. We use the orthonormal frame $\hat\theta$ and $\hat q$. Although the tangent direction is obvious in this example, the following definition generalizes directly: a variation $\delta q_\theta$ is tangent to the constraint surface if it preserves the constraint to first order. Linearizing $C(q)$ gives
\begin{equation}
    \delta C(q) = \nabla C(q) \cdot\delta q_\theta= 2 q\cdot\delta q_\theta=0\, .
    \label{eq:circle-linearized-constraint}
\end{equation}

On shell of the constraint, the equation of motion decomposes as
\begin{equation}
    \label{eq:circle-constrained-eom-decomposed}
    \left(m\ddot q+\nabla U(q) - \lambda q\right)\cdot \delta q_\theta +\left(m\ddot q+\nabla U(q) - \lambda R\hat q\right) \cdot \delta q_n =0 \, .
\end{equation}
We use the radial equation to fix $\lambda$ in terms of the radial acceleration $\ddot q_\rho=\ddot q \cdot \hat q$ and the radial component of the potential gradient $\partial_\rho U(q) = \nabla U(q) \cdot \hat{q}$:
\begin{equation}
    \label{eq:circle-normal-eom}
    m\ddot q_\rho+\partial_\rho U(q)-\lambda R=0\, , \quad \longrightarrow \quad \lambda = \frac{m\ddot q_\rho+\partial_\rho U(q)}{R}\, .
\end{equation}
The radial equation carries no independent physical content; it only fixes the value of the Lagrange multiplier. Projecting along the tangent direction gives
\begin{equation}
    \label{eq:circle-tangent-eom}
    \left(m\ddot q+\nabla U(q) - \lambda q\right)\cdot \delta q_\theta = 0\, .
\end{equation}
By \eqref{eq:circle-linearized-constraint}, tangentiality means $q \cdot \delta q_\theta=0$. The Lagrange multiplier $\lambda$ therefore drops out of the tangential equation, confirming that it is fixed only by the radial equation. Denoting by $\ddot q_\theta = \ddot q \cdot \hat{\theta}$ the tangential acceleration, and by $\partial_\theta U(q) = \nabla U(q) \cdot \hat{\theta}$ the tangential component of the potential gradient, the tangential equation becomes
\begin{equation}
    m\ddot q_\theta + \partial_\theta U(q) = 0\, .
\end{equation}
This is exactly the intrinsic equation of motion \eqref{eq:circle-intrinsic-eom} once we identify $\ddot q_\theta=R\ddot\theta$ and $\partial_\theta U(q)=\frac{1}{R}\frac{\d U_R}{\d\theta}$. Indeed, differentiating $\dot q=\dot\theta\,q'(\theta)$ once more and using $q''(\theta)=-q(\theta)$ gives $\ddot q=R\ddot\theta\,\hat\theta-R\dot\theta^2\,\hat q$, so that $\ddot q_\theta=R\ddot\theta$; and the chain rule gives $\nabla U(q)\cdot q'(\theta)=\d U_R/\d\theta$, so that $\partial_\theta U(q)=\frac{1}{R}\d U_R/\d\theta$. Multiplying the tangential equation by $R$ then reproduces \eqref{eq:circle-intrinsic-eom} exactly.

Conversely, let $\theta(t)$ solve \eqref{eq:circle-intrinsic-eom} and set $q(t) \equiv q(\theta(t))$. Then $q^2=R^2$ holds identically. Dividing \eqref{eq:circle-intrinsic-eom} by $R$ and using the identification above shows that $m\ddot q_\theta+\partial_\theta U(q)=0$, i.e.\ that the tangential component of $m\ddot q+\nabla U(q)$ vanishes, $\hat\theta\cdot\bigl(m\ddot q+\nabla U(q)\bigr)=0$. Since $\hat\theta$ and $\hat q$ span the plane, $m\ddot q+\nabla U(q)$ is then purely radial,
\begin{equation}
    m\ddot q+\nabla U(q)=\lambda q\, ,
    \qquad
    \lambda \equiv \frac{q\cdot\bigl(m\ddot q+\nabla U(q)\bigr)}{R^2}\, ,
\end{equation}
which is exactly the $q$-equation appearing in \eqref{eq:circle-constrained-eom}, with $\lambda$ fixed as in \eqref{eq:circle-normal-eom}. So every intrinsic solution lifts to a constrained solution. The value of $\lambda$ plays no role in selecting the physical trajectory.

What do we learn from this simple example? Checking that the constrained system describes the same physics as the intrinsic one does not require solving either set of equations of motion explicitly: 
\begin{itemize}
    \item It is enough to check that the tangential part of the ambient equation of motion, \eqref{eq:circle-tangent-eom}, reproduces the intrinsic equation of motion \eqref{eq:circle-intrinsic-eom}. In general, the tangential part of the ambient equation of motion is obtained by projecting the ambient equation along the tangent space of the constraint surface. The Lagrange multipliers drop out of the tangential equation because they are normal to the constraint surface.
    \item The radial part of the ambient equation of motion, \eqref{eq:circle-normal-eom}, determines the Lagrange multiplier $\lambda$. This is also a general feature: the normal part of the ambient equation of motion fixes the Lagrange multipliers, but does not constrain the physical trajectory.
\end{itemize}

\bibliographystyle{unsrt}
\bibliography{biblio}

\end{document}